\documentclass[10pt]{revtex4}
\usepackage[dvipdfm]{graphicx}
\usepackage{amsmath,amssymb,amsthm,bm}

\theoremstyle{plain}
\newtheorem{thm}{Theorem}
\newtheorem{lem}{Lemma}

\newtheorem{prop}{Proposition}

\newcommand{\balpha}{\bm{\alpha}}
\newcommand{\bal}{\bm{\alpha}}
\newcommand{\bbeta}{\bm{\beta}}
\newcommand{\bep}{\bm{\epsilon}}
\newcommand{\bet}{\bm{\eta}}
\newcommand{\teta}{\widetilde\eta}
\newcommand{\tk}{\widetilde k}

\newcommand{\tq}{\widetilde q}
\newcommand{\tbet}{\bm{\widetilde\eta}}
\newcommand{\eps}{\epsilon}
\newcommand{\vareps}{\varepsilon}
\newcommand{\defi}{\stackrel{\rm def}{=}}
\newcommand{\defeq}{\stackrel{\rm def}{=}}
\newcommand{\be}{\begin{equation}}
\newcommand{\ee}{\end{equation}}
\newcommand{\bp}{\begin{proposition}}
\newcommand{\ep}{\end{proposition}}
\newcommand{\pro}{\begin{proof}}
\newcommand{\epro}{\end{proof}}
\newcommand{\U}{\widetilde{U}}
\newcommand{\tF}{\widetilde{F}}
\newcommand{\tC}{\widetilde{C}}
\newcommand{\tcU}{\widetilde{\mathcal{U}}}
\newcommand{\cC}{\mathcal{C}}
\newcommand{\cL}{\mathcal{L}}
\newcommand{\cU}{\mathcal{U}}
\newcommand{\cO}{\mathcal{O}}
\newcommand{\cN}{\mathcal{N}}
\newcommand{\cS}{\mathcal{S}}
\newcommand{\IZ}{\mathbb{Z}}
\newcommand{\IR}{\mathbb{R}}
\newcommand{\IC}{\mathbb{C}}
\newcommand{\IN}{\mathbb{N}}
\newcommand{\Can}{{\rm Can}}
\newcommand{\ra}{\rangle}
\newcommand{\la}{\langle}
\newcommand{\supp}{\operatorname{supp}}
\newcommand{\vis}{v}
\newcommand{\reshuff}{\mathrm{resh}}

\begin{document}
\title{On the resonance eigenstates of an open quantum baker map}
\author{J.P. Keating$^1$, S. Nonnenmacher$^2$, M. Novaes$^1$ and M. Sieber$^1$}
\affiliation{$^1$School of Mathematics, University of Bristol,
Bristol BS8 1TW, United Kingdom\\$^2$Institut de Physique Th\'eorique,
CEA, IPhT, F-91191 Gif-sur-Yvette, France;
CNRS, URA 2306, F-91191 Gif-sur-Yvette, France}

\begin{abstract}
We study the resonance eigenstates of a particular quantization of the open
baker map. For any admissible value of Planck's constant, the corresponding
quantum map is a subunitary matrix, and the nonzero component of its spectrum
is contained inside an annulus in the complex plane, $|z_{min}|\leq |z|\leq
|z_{max}|$. We consider semiclassical sequences of eigenstates, such that the
moduli of their eigenvalues converge to a fixed radius $r$. We prove that, if
the moduli converge to $r=|z_{max}|$, then the sequence of eigenstates
is associated with a fixed phase space measure $\rho_{max}$. The same holds for
sequences with eigenvalue moduli converging to $|z_{min}|$, with a different
limit measure $\rho_{min}$. Both these limiting measures are supported on
fractal sets, which are trapped sets of the classical dynamics. For a general
radius $|z_{min}|\leq r \leq |z_{max}|$ there is no unique limit measure, and
we identify some families of eigenstates with precise self-similar properties.

\end{abstract}

\maketitle

\section{Introduction}
In the semiclassical limit, stationary states of quantum systems are in general
expected to concentrate on phase space structures that are invariant under the
corresponding classical dynamics \cite{berry,voros}. More precisely, any
semiclassical sequence of eigenstates of energies $\sim E$ is associated with
one or several semiclassical limit measures, which are probability measures on
the energy shell $\Sigma_E$, invariant under the Hamiltonian flow. We will say
that the semiclassical sequence of eigenstates converges to one or several
limit measures.

According to the Quantum Ergodicity theorem
\cite{ergod,cmp102ycv1985,jpa20sz1987,helffer}, if the classical
flow on some energy shell $\Sigma_E$ is ergodic with respect to the
Liouville measure, then in the semiclassical limit almost all
quantum eigenstates of energy $\sim E$ become (in a weak sense)
equidistributed on the energy shell; for a recent review, see for
instance \cite{zelditch}. It means that almost all such
semiclassical sequences converge to the Liouville measure on
$\Sigma_E$.

Canonical maps on the two-dimensional torus, such as the baker map
\cite{baker1, baker2, baker3}, the cat map \cite{Hannay, k1} and their
generalizations \cite{kmr}, can be quantized to give finite-dimensional unitary
matrices. Such quantum maps have been widely used as toy models in the context
of quantum chaos, because they possess several simplifying properties:
two-dimensional torus phase space instead of $\Sigma_E$, simple symbolic
dynamics, finite dimensional quantum mechanics, easy numerical implementation
(see, e.g., \cite{mirkobook}).  If the map is ergodic with respect to the
Liouville (=Lebesgue) measure, then a Quantum Ergodicity theorem applies that
almost all sequences of eigenstates of the quantized map become uniformly
distributed on the torus in the semiclassical limit
\cite{cmp178ab1996,esposti,rudnick,rudnick07}.  (Quantum ergodicity has also
recently been established for certain families of quantum graphs \cite{BKS},
but for other families it is known not to hold \cite{star1,star2}.)

It is important to remark that, in general, a chaotic system admits
many invariant measures different from the Liouville measure, e.g.
delta measures carried on periodic orbits, or fractal measures. Even
when the system is quantum ergodic, exceptional sequences of
eigenstates could converge to some of these invariant measures. Such
exceptional eigenstates have been constructed for several types of
quantized ergodic maps on the torus
\cite{schubert,cmp239ff2003,entropy} and for certain quantum graphs
\cite{star1,star2}. On the other hand, ergodic systems for which all
sequences of eigenstates converge to the Liouville measure are said
to obey Quantum Unique Ergodicity (QUE). This is obviously the case when
there exist no invariant measures other than the Liouville measure
\cite{gfa10jm2000,lior}. Some ergodic systems carry arithmetic
structures (typically, a commuting family of ``Hecke'' operators
commuting with the quantum dynamics), which lead one to consider
joint (``Hecke'') eigenstates. In that case, one can sometimes prove
that all the Hecke eigenstates semiclassically converge to the
Liouville measure, a property called Arithmetic Quantum Unique
Ergodicity \cite{am163el2006,dmj103pk2005,cmp161zr2005}.

The corresponding properties of open (scattering) chaotic systems are currently
not nearly as well understood. The invariant properties of the classical open
system are related to the set of trapped trajectories, called the trapped set.
For a chaotic system, this trapped set is generally a {\em fractal repeller}.
The statistical properties of the system can be associated with conditionally
invariant measures, or eigenmeasures of the propagator, which are only
invariant up to a constant (the decay rate of the measure). A recent review of
the theory of classical chaotic systems with openings can be found in
\cite{n19mfd2006}. In the quantum version of the problem each resonant state
has a specific decay rate. For the system studied here we show that a form of
QUE holds for {\it extremal} decay rates, whilst multiple limit measures exist
in the bulk. Because the notion of ergodicity is not clearly defined
for the open system, it would be more correct to use the phrase ``quantum uniqueness''
rather than QUE, but the latter has the advantage to connect us with previous works
on closed chaotic systems.

\subsection{Open chaotic maps}

Instead of dealing with a genuine scattering system (with infinite-volume phase
space), it is simpler to consider a discrete-time dynamical system (that is, a map)
living on a compact phase space, but with a ``hole'' through which the particles escape, never to
return (the map is then said to be ``open'').
This compact phase space is a model for the ``interaction region'' of a
realistic scattering system. Open maps may also be quantized, into
subunitary matrices (see below). One expects the spectrum of these quantum maps
to (at least qualitatively) mimic the properties of the resonances and resonant
states of scattering systems.

We focus here on a specific quantization of the open baker map, for
which we prove some conjectures pertaining to the semiclassical
behaviour of (resonant) eigenstates \cite{previous,Rubin}. Before
stating our results, we describe the general construction of an open
map on the torus.

Consider an invertible canonical map $\cU$ on the 2-dimensional torus $T$
(viewed as a phase space). We assume that $\cU$ is chaotic, in particular that
it is ergodic with respect to the Liouville measure on $T$. Let this map be
``opened'' by identifying some region of phase space with a ``hole'' through
which points escape, and denote the resulting open map by $\tcU$. The
ergodicity of $\cU$ implies that almost all initial conditions escape the
system when propagated either forwards or backwards. The set of initial
conditions that remain trapped for infinite times when propagated to the future
(respectively the past) is called the forward-trapped (respectively
backward-trapped) set and denoted by $K_-$ (respectively $K_+$).

All invariant measures are supported on the intersection $K_0=K_+\cap K_-$,
which is the trapped set or the repeller. The sets $K_-$ and $K_+$ are the
stable and unstable manifolds of the repeller, respectively. We denote the
opening by $O$, and by $O_m=\cU^m(O)$ its image under the $m$th power of
the closed map. The trapped sets are then defined as \be
K_-=T\setminus\bigcup_{m=0}^\infty O_{-m},\quad
K_+=T\setminus\bigcup_{m=1}^\infty O_{m}. \ee

\subsection{Open quantum maps}
We choose to quantize the torus using {\em antiperiodic} boundary
conditions for the wavefunctions, so that both position and momentum
take values of the form $({\rm integer}+1/2)/N$ \cite{Hannay}. In
the position representation, quantum states are thus given by a
``comb'' of delta functions supported on the set \be\label{sn}
S_N=\left\{q_j=\frac{j+1/2}{N},\ \ 0\leq j\leq N-1\right\}\,. \ee
These quantum states form a Hilbert space of dimension $N$, where
$N$ plays the role of an effective Planck's constant, $\hbar=(2\pi
N)^{-1}$. The quantization of the invertible canonical map $\cU$ is
a unitary matrix $U$ acting on this Hilbert space. Its eigenvalues
therefore lie on the unit circle.

We shall open our map by taking the hole to be a strip parallel to the momentum
axis, of the form $O=I\times [0,1)$. At the quantum level, we call $\Pi$ the
projector on the subspace spanned by the positions in $q_j\in I$. The
complementary projector
$(1-\Pi)$ kills (at each step of the dynamics) the states localized in the
hole, and let the others evolve through the propagator $U$. Hence, the quantum
version of the open map is given by the (nonunitary) matrix $\U=U(1-\Pi)$. The
result of multiplying by $1-\Pi$ is to set to zero the columns of $U$
corresponding to the hole. The matrix $\U$ is not normal, so we must
distinguish between its right and left eigenstates, \be\label{leftright}
\U|\Psi^R_n\ra=z_n|\Psi^R_n\ra, \quad \la \Psi_n^L|\U=z_n\la \Psi_n^L|. \ee We
shall assume the eigenstates to be normalized according to $\la
\Psi_n^L|\Psi^L_n\ra=\la \Psi_n^R|\Psi^R_n\ra=1$, instead of the usual $\la
\Psi_n^L|\Psi^R_m\ra=\delta_{nm}$. The eigenvalues $z_n$ of $\U$ lie in the
unit disk, $|z_n|^2=e^{-\Gamma_n}\leq 1$; $\Gamma_n\geq 0$ is called the decay
rate associated with the eigenstate $|\Psi^R_n\ra$.

The usual Weyl law for closed systems relates (in the semiclassical limit) the
mean eigenvalue density to the available phase space volume. For open chaotic
systems the mean density of resonances is believed to be determined by the
fractal dimension of the repeller, a property known as the fractal Weyl law
\cite{prl91wl2003}. This law was investigated numerically for two-dimensional
Hamiltonian scattering systems \cite{cpl201kkl2002,glz2004,prl91wl2003}, for
the baker map \cite{jpa38sn2005,cmp269sn2007} and for the kicked rotator
\cite{henning}. It has been proven for the Walsh-baker map
\cite{jpa38sn2005,cmp269sn2007} (see below). Although a fractal upper bound on
the number of resonances has been proven for Hamiltonian scattering systems
\cite{frac2}, the fractal Weyl law still stands as a conjecture for generic
open systems. In \cite{henning} a heuristic argument for this law was
presented, based on the distinction between {\it short-lived} and {\it
long-lived} eigenstates. The short-lived states are associated with phase space
regions that escape in a short time (comparable to the Ehrenfest time), so that
their decay rate satisfies $\Gamma_n\gg 1$. On the other hand, long-lived
states remain in the system long enough to develop strong diffraction and
interference effects, leading to a finite decay rate $\Gamma_n=\cO(1)$. The
latter will be the main focus of our investigation.

\subsection{Semiclassical limit of ``resonant'' eigenstates}
To further motivate the present work, we here summarize some recent heuristic
arguments and numerical results. Eigenstates of open chaotic maps were studied
in a general setting by Keating {\it et al} in \cite{previous}. They were
represented in phase space by using Husimi functions \be\label{hdef}
H_\psi(x)=\frac{1}{2\pi \hbar}|\la x|\psi\ra|^2,\ee where $x$ is a point in
phase space and $|x\ra$ a standard Gaussian coherent state. It was shown that,
in the semiclassical limit, the phase space support of long-lived right
(respectively left) eigenstates must be a subset of the backward (respectively
forward) trapped set, in the sense that \be\label{limit} \lim_{\hbar\to 0}
H_{\Psi^R_n}(x)= 0 \quad {\rm if }\;x\notin K_+\,,\qquad\text{resp.}\quad
\lim_{\hbar\to 0}H_{\Psi^L_n}(x)= 0 \quad {\rm if }\;x\notin K_-\,. \ee For
maps on the torus it was shown that if $\Pi_1=\Pi$ is the projector onto the
opening and $\Pi_{m+1}$ the projector onto the set of points which reach the
opening after $m$ steps but not earlier, then within the semiclassical
approximation one has $\Pi_{m+1}\approx
(\widetilde{U}^\dag)^m\Pi_1\widetilde{U}^m$ and therefore \be\label{weights}
\la\Psi^R_n|\Pi_{m+1}|\Psi^R_n\ra\approx|z_n|^{2m}(1-|z_n|^2). \ee

\begin{figure}[t]
\includegraphics[scale=0.6]{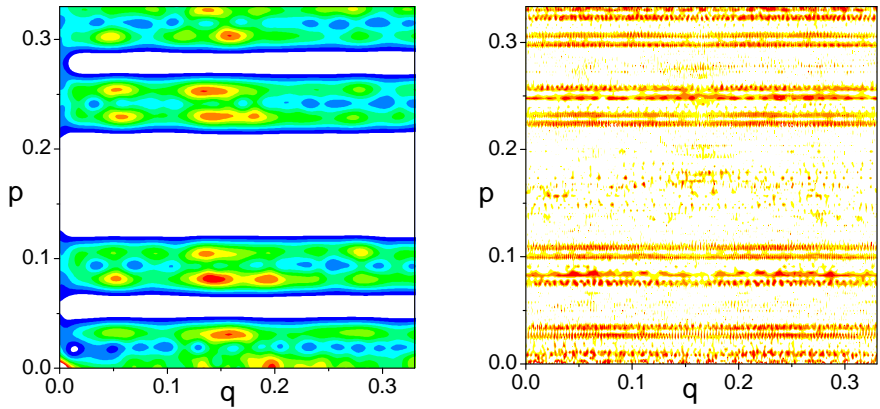}
\caption{(color online) Left panel: The average of the Husimi
functions (\ref{hdef}) of the $100$ longest-lived right eigenstates
of the baker map, for $N=3^7$ (intensity increases from blue to
red). Right panel: the corresponding Wigner function average (in the
white regions the function is non-positive). Taken from \cite{previous}.\label{fig1}}
\end{figure}

The specific system studied numerically in \cite{previous}
was the triadic baker map, defined by
\be\label{classical}
\mathcal{U}(q,p)=\begin{cases}(3q,\frac{p}{3}) &\text{if } 0\leq q<\frac{1}{3}, \\
(3q-1,\frac{p+1}{3}) &\text{if } \frac{1}{3}\leq
q<\frac{2}{3},\\(3q-2,\frac{p+2}{3}) &\text{if } \frac{2}{3}\leq
q<1.\end{cases} \ee This map was opened by sending ``to infinity'' points in
the middle vertical strip; in this case, $K_-$ is the product
$\Can\times[0,1)$, where $\Can$ denotes the usual middle-third Cantor set, and
$K_+=[0,1)\times\Can$. The closed baker map was quantized in
\cite{baker1,baker2,baker3}.  Taking the opening into account one obtains the
matrix \cite{sv1996} \be\label{quantumbaker}
\U=F_N^\dag\left(\begin{array}{ccc}F_{N/3} & &   \\& 0 &   \\&   & F_{N/3}
\\\end{array}\right)\,. \ee Here $F_N$ is the discrete Fourier transform on the
set $S_N$, \be\label{fourier} (F_N)_{nm}=\frac1{\sqrt{N}}e^{-\frac{2\pi
i}{N}(n+1/2)(m+1/2)},\quad n,m\in\{0,\ldots,N-1\}. \ee The numerical
computations in \cite{previous} were restricted to values of
$N=(2\pi\hbar)^{-1}$ given by $N=3^{k}$, such that the semiclassical limit
corresponds to $k\to \infty$. In Figure~\ref{fig1} we plot the Husimi function
of right eigenstates, averaged over the $100$ longest-living states, for the
case $N=3^7$. This function is approximately concentrated on the
backward-trapped set. The right panel shows the averaged Wigner function
\cite{Hannay}, which resolves $K_+$ with higher accuracy.

Nonnenmacher and Rubin considered eigenstates of the open quantum
baker \eqref{quantumbaker} in \cite{Rubin} (in their case, $F_N$ is
the discrete Fourier transform on the set $\{j/N,\
j=0,\ldots,N-1\}$). They showed, in particular, that if a point $x$
is at a finite distance from $K_+$ and from the set of
discontinuities of the classical map $\cU$, then in the
semiclassical limit the property \eqref{limit} is a consequence of
the stronger estimate \be H_{\Psi^R_n}(x)=O(e^{-cN}),\quad
N\to\infty\,. \ee They also showed that for any semiclassical
sequence of (right) eigenstates $(\Psi_N)_{N\to\infty}$ such that
$\lim_{N\to\infty}|z_N|^2=e^{-\Gamma}$, the Husimi measures
$H_{\Psi_N}$ weak-$*$ converge to one or several probability
measures on the torus: each of these limit measures is necessarily
an {\em eigenmeasure} (also called conditionally invariant measure)
of the open map $\tcU$, with the eigenvalue $e^{-\Gamma}$. The
semiclassical estimate \eqref{weights} is then a consequence of the
conditional invariance of limit measures.

\subsection{The Walsh-baker map}\label{s:Walsh-baker}
When $N=3^k$, an alternative quantization of the baker map
\eqref{classical} exists, which is based on a modified Fourier
transform, the Walsh-Fourier transform
\cite{jpa38sn2005,cmp269sn2007}. The Walsh-baker map can be solved
exactly, with explicit expressions for the spectrum and the
eigenstates. Its open version is the only system for which a fractal
Weyl law has been rigorously proven. In the Walsh framework, the
phase space distribution of eigenstates can be studied through a
Walsh version of the Husimi measure, called the Walsh-Husimi measure
(see section~\ref{s:WalshCS}). This measure is defined on ``quantum
rectangles'' (also called ``tiles'' \cite{thiele}).

Let us recall the definition of a ``rectangle'' in phase space. Let us denote
by \be q-\frac1{2N}=0\cdot\epsilon_1\epsilon_2\epsilon_3\cdots\ee the number
$q-\frac1{2N}\in[0,1)$, the ternary decomposition of which is given by \be
q-\frac1{2N}=\sum_{\ell=1}^{\infty}\epsilon_\ell\, 3^{-\ell}, \quad
\epsilon_\ell\in\{0,1,2\}\,. \ee For any $b\in\IN$, a sequence
$\bep=\eps_1\cdots\eps_b$ describes an interval of length $3^{-b}$ in the
position axis, which we denote by $[\bep]$. If, for some $b'\in\IN$,
$[\bep'=\eps'_1\cdots\eps'_{b'}]$ describes a similar triadic interval in the
momentum axis, then the product of these two intervals, which is a rectangle of
area $3^{-b-b'}$, will be denoted by $[\eps'_{b'}\cdots\eps'_1\cdot
\eps_1\cdots\eps_b]$ (notice the inversion of indices for $\bep'$).

In the following, we will often refer to the case where $b=b'=\vis$, for some
fixed $\vis>0$, which we will call a {\it $\vis$-square} and denote it by
$[\bep'\cdot\bep]_{\vis}$. A rectangle of size $v\times 0$ will be called {\it
vertical}, and denoted by $[\cdot\bep]_{\vis}$. If on the other hand $b,b'$ are
related by $b'=k-b$ (where $k$ is the quantum parameter), then
$[\bep'\cdot\bep]$ will be called a ``quantum rectangle''. For each $b\in
[0,k]$, the Walsh-Husimi measure $H^b_{\Psi}$ is defined by its weights on the
rectangles of size $b\times (k-b)$, denoted by $H^b_{\Psi}([\bep'\cdot\bep])$.
One recovers the position (respectively Walsh-momentum) representation by
taking $b=k$ (respectively $b=0$):
\begin{align}
H^k_{\Psi}([\cdot\bep])&=|\la q|\Psi\ra|^2 \quad
\text{for}\quad q=0.\eps_1\cdots\eps_k+\frac1{2N}\,\\
H^0_{\Psi}([\bep'\cdot])&=|\la p|\Psi\ra|^2 \quad
\text{for}\quad p=0.\eps'_1\cdots\eps'_k+\frac1{2N}\,.
\end{align} We will need to compute the value of a Walsh-Husimi measure on `classical'
$\vis$-squares $[\bep'\cdot\bep]_{\vis}$, for which $\vis$ is independent of
$k$. We define this in the natural way, taking
$H^{b}_{\Psi}([\bep'\cdot\bep]_{\vis})$ to be the sum of the values of
$H^{b}_{\Psi}$ over all quantum squares contained in $[\bep'\cdot\bep]_{\vis}$
(see (\ref{Honvquare})).

The Walsh quantization of the {\em closed} baker map was studied in
\cite{entropy}: the authors proved quantum ergodicity but found
several counterexamples to quantum unique ergodicity. In particular,
they constructed semiclassical sequences of eigenstates along which
the Walsh-Husimi measures converge to (multi)fractal invariant
measures.

In the present work we focus on the Walsh quantization of the open
baker, defined in \cite{jpa38sn2005,cmp269sn2007} and further
studied in \cite{Rubin}. The quantum propagator is given by the
following matrix ($N=3^k$): \be\label{WalshB}
B=W_N^\dag\left(\begin{array}{ccc} W_{N/3} &   &
\\  & 0 &   \\  & & W_{N/3} \\\end{array} \right),
\ee where $W_N$ is the Walsh-Fourier transform (see \eqref{walsh}).
The non-zero part of the spectrum forms a lattice inside the annulus
$\{|z_{min}|\le |z|\le|z_{max}|\}$. We use a quantization slightly
different from the one used in \cite{Rubin}. In our case the
eigenvalues at the ``edges'' of the nontrivial spectrum are given by
$$
z_{max}=1\quad\text{and}\quad z_{min}=\frac{i}{\sqrt{3}}\,,
$$
both being non-degenerate. On the other hand, eigenvalues in the
``bulk'' of the spectrum $\{|z_{min}| < |z| < |z_{max}|\}$ are (highly)
degenerate. The kinematics of the map is such that the position
representation of right eigenstates is equal to the (Walsh-)momentum
representation of left eigenstates. We may therefore restrict our
analysis to the study of the right eigenstates.

\subsection{Statement of our results}\label{s:statement}

We obtain precise results on the phase space distribution of long-living
eigenstates of the open Walsh-baker map, both for finite $N=3^k$ and in the
limit $k\to\infty$. One question posed in \cite{Rubin} concerns the family of
eigenmeasures one can obtain by taking weak-$*$ semiclassical limits of Husimi
(or Walsh-Husimi) measures: for a given $\Gamma\geq 0$, which eigenmeasures of
eigenvalue $e^{-\Gamma}$ can be obtained as semiclassical limits? It was
conjectured that the (right) eigenstates of $B$ near the ``edges'' of the
nontrivial spectrum, i.e. such that $|z|\to |z_{max}|$ (resp. $|z|\to
|z_{min}|$), converge to a unique measure $\rho_{max}$ (resp. $\rho_{min}$). In
the theorem below we prove this conjecture.

An important role is played here by the Cantor set and its generations. If we
denote by $\Can_n$ the set of points in the interval $[0,1)$ such that the
first $n$ digits in their ternary decomposition are all different from $1$, the
sequence $(\Can_{n})_{n\geq 1}$ converges to the middle-third Cantor set
$\Can=\Can_\infty$. The ``uniform measure'' on $\Can$ is defined as follows on
triadic intervals of length $n$: \be\label{e:nu_Can} \forall
\bep=\eps_1\cdots\eps_n\,,\qquad \nu_{\Can}([\bep])\defeq
\begin{cases}
2^{-n}\quad \text{if all }\eps_i\in\{0,2\}\\
0\quad \text{otherwise.}
\end{cases}
\ee
\begin{thm}\label{th:1}
Let $(\Psi_k)_{k\to\infty}$ be a sequence of right eigenstates of the open
Walsh-baker $B$, such that the associated eigenvalues semiclassically converge
to the outer edge of the spectrum: $|z_k|\stackrel{k\to\infty}{\longrightarrow}
1$. Then the corresponding Husimi measures $(H^{[k/2]}_{\Psi_k})$ weak-$*$
converge to a unique invariant measure $\rho_{max}$,
supported on the repeller $K_0$. This means that
$H^{[k/2]}_{\Psi_k}([\bep'\cdot\bep]_{\vis})
\stackrel{k\to\infty}{\longrightarrow}\rho_{max}([\bep'\cdot\bep]_{\vis})$
for any $\vis$-square $[\bep'\cdot\bep]_{\vis}$.

Similarly, if the eigenvalues of a sequence $(\Psi_k)_{k\to\infty}$ of right
eigenstates converge to the inner edge, $|z_k|\to 1/\sqrt{3}$, then the Husimi
measures $H^{[k/2]}_{\Psi_k}$ converge (in the above sense) to a unique
self-similar eigenmeasure $\rho_{min}$ supported on $K_+$.

Both measures $\rho_{max}$ and $\rho_{min}$ can be factorized as
$\rho=\nu_{\Can}(dp)\times\nu_{max/min}(dq)$, where $\nu_{\Can}$ is the uniform
measure \eqref{e:nu_Can} on the Cantor set, $\nu_{max}=\nu_{\Can}$ and
$\nu_{min}$ is a certain self-similar measure on $[0,1)$ (see \eqref{e:numin}).
\end{thm}
$\rho_{max}$ is an invariant measure of the (closed) baker map $\cU$, localized
on the trapped set $K_0 $. Loosely speaking, it is the ``uniform'' measure on
$K_0$. More precisely, it is the invariant measure of maximal entropy (and at
the same time the Gibbs measure associated with the potential $-\log J^u(x)$)
for the restriction of $\cU$ to $K_0$ (see \cite{CherMar97} for the description
of these measures in a more general context).

Theorem 1 expresses a form of ``quantum uniqueness'' at
the edges of the nontrivial spectrum of $B$. The
next theorems apply to the ``bulk'' of the nontrivial spectrum, and show
that such a quantum uniqueness does not hold there. These theorems are
concerned with a particular eigenbasis for the nontrivial spectrum (see
\S\ref{s:eigenstates}).

The description of this particular eigenbasis uses binary sequences
$\bet=\eta_1\cdots \eta_k$ of length $k$, such that each symbol
$\eta_j\in\{+,-\}$. If $\bet$ is a periodic sequence of period $\ell$, its
orbit under the cyclic shift (which we denote by $[\bet]$) consists of $\ell$
different sequences. We can associate with this orbit $\ell$ different
long-lived right eigenstates of $B$, denoted by $\Psi^{m}_{\bet}$ (see
\eqref{e:Psi-dm} in \S\ref{s:eigenstates}). The respective eigenvalues are
$z_{\delta,m/\ell}=z_{min}^{\delta}\,e^{2\pi im/\ell}$, where $0\leq
m\leq\ell-1$. The number of times the symbol `+' appears in $\bet$ is called
its {\em degree} and denoted by $d$. The variable $\delta\in [0,1)$ is given by
$\delta=d/k$ and called the {\em relative degree} of $\bet$.

For any $0\leq b\leq k$, we denote by $H^b_{\Psi^{m}_{\bet}}$ the Walsh-Husimi measure
associated with this eigenstate. Before dealing with these individual eigenstates,
it is easier to average over the phase index $m$, and define the averaged
Husimi measure
\be\label{e:average-Hus}
H^b_{[\bet]}\defi
\frac{1}{\ell}\sum_{m=0}^{\ell-1} H^b_{\Psi^{m}_{\bet}}\,.
\ee
Our understanding of these averaged measures is not only semiclassical, but already valid
for finite $k$.
\begin{thm}\label{th:2}
For any value of $k\geq 1$, any $0\leq b\leq k$, and any sequence
$\bet$ of length $k$ and degree $d=\delta k$, the averaged
Walsh-Husimi measure $H^b_{[\bet]}$ is equal to a certain
eigenmeasure $\rho_{[\bet]}$ with eigenvalue
$e^{-\Gamma}=3^{-\delta}$, conditioned on the rectangles of size
$b\times(k-b)$. This measure is of the form \be\label{e:rho_bet}
\rho_{[\bet]}=\nu_{\Can}(dp)\times\nu_{[\bet]}(dq)\,, \ee where
$\nu_{\Can}$ is the uniform measure \eqref{e:nu_Can} on $\Can$, and
$\nu_{[\bet]}$ is a certain probability measure on the interval.
$\nu_{[\bet]}$ satisfies the following self-similarity properties:

i) for any $0\leq n\leq b$ and any sequence $\eps_1\cdots\eps_b$ such that
$\eps_1,\eps_2,\ldots,\eps_n\in\{0,2\}$, one has
\be\label{e:scaling}
\nu_{[\bet]}([\eps_1\cdots\eps_n\cdots\eps_b])=(2\times 3^{\delta})^{-n}\;
\nu_{[\bet]}([\eps_{n+1}\cdots\eps_b])\,.
\ee

ii) In particular, for any sequence $\bep\in\{0,2\}^n$, the interval
$[\bep]\subset \Can_n$ has weight $\nu_{[\bet]}([\bep])=(2\times
3^{\delta})^{-n}$. In general, for any $b$-sequence $\bep$
containing $n$ symbols $\eps_i\neq 1$, we have the upper bound
\be\label{e:bounds} \nu_{[\bet]}([\bep])\leq
\frac1{2^{n}}\,\big(\frac23\big)^{b-n}\,. \ee
\end{thm}
The last bound, together with the definition of $\nu_{\Can}$,
restricts the concentration of $\rho_{[\bet]}$.

\begin{figure}[ht]
\begin{center}
\includegraphics[scale=0.6]{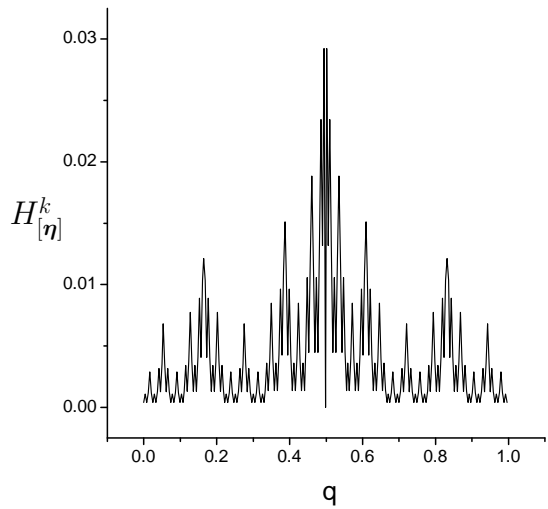}
\end{center}
\caption{The position density function $H^k_{[\bet]}(q)$ for
${\bet}=++++-$.\label{f:fig3}}
\end{figure}

Adapting \cite[Prop.8]{Rubin} to the present choice of Walsh quantization, we
see that for any fixed primitive sequence $\bet_0$ of length $k_0$, the measure
$\rho_{[\bet_0]}$ is the semiclassical limit of the sequence of eigenstates
$\big(\Psi^m_{(\bet_0)^n}\big)_{n\to\infty}$, where $m\in\{0,\ldots,k_0-1\}$
can vary arbitrarily with $n$. The above theorem shows that $\rho_{[\bet_0]}$
already coincides, for {\em finite} $k_0$, with the averaged Husimi measure
\eqref{e:average-Hus}. In Figure~\ref{f:fig3} we present as an example the
spectral average $H^k_{[\bet]}$ for $N=3^5$ and ${\bet}=++++-$ (equivalently,
the measure $\nu_{[\bet]}$ conditioned on the intervals
$[\eps_1\cdots\eps_5]$).

The next result, which requires much more effort,
shows that the measures $\rho_{[\bet]}$ are also
relevant to describe the {\it individual} Husimi measures $H^b_{\Psi^{m}_{\bet}}$.
\begin{thm}\label{th:3}
Fix a $\vis$-square $[\bep'\cdot\bep]_{\vis}$. Take any sequence $\bet$ of
length $k$, any $m\in\{0,\ldots,k-1\}$ and any $b\geq\vis$ such that $k-b\ge
v$. Then, \be\label{e:stat-indiv}
H^{b}_{\Psi^{m}_{\bet}}([\bep'\cdot\bep]_{\vis}) =
\rho_{[\bet]}([\bep'\cdot\bep]_{\vis}) + \cO_{\vis}(k^{-1}),\qquad
k\to\infty\,, \ee where $\rho_{[\bet]}=\nu_{\Can}\times\nu_{[\bet]}$ is the
eigenmeasure described in Theorem~\ref{th:2}. The implied constant is
independent of $\bet$, $m$ and $b$.
\end{thm}
When considering sequences $\bet$ of lengths $k\to\infty$, any
weak-$*$ limit of the measures $(\rho_{[\bet]})$ will be of the form
$\nu_{\Can}(dp)\times \nu(dq)$ for some probability measure $\nu$ on
the unit interval. So far all the semiclassical measures we have encountered
are of that form. One might wonder whether this is the case for {\it all} semiclassical
measures of the Walsh-baker \eqref{WalshB}. Our
last theorem answers this question in the negative.
\begin{thm}\label{th:4}
For any $z\in\IC$ in the bulk of the nontrivial spectrum (that is, $1/\sqrt{3}<|z|<1$),
there exists an explicit sequence of
eigenstates $(\Psi_k)_{k\to\infty}$ with eigenvalues $z_k\to z$, converging to
a semiclassical measure $\rho$ which is not of the form $\nu_{\Can}(dp)\times \nu(dq)$.
\end{thm}

\noindent{\bf Remark.} In ref.\cite{cmp269sn2007} it was noticed that the matrix
\eqref{WalshB} can also be interpreted as the ``standard'' (Weyl-like)
quantization of a multivalued symplectic map constructed
``above'' the baker's map \eqref{classical}. Within this interpretation,
the phase space properties of the eigenstates should
be analyzed through the {\em standard} Husimi measures \eqref{hdef} instead of the Walsh-Husimi ones.
These two measures generally differ in the vertical (momentum)
direction, but their projections $\pi_*\rho$ on the position axis are similar to each other.
As a result, for a given sequence of eigenstates $(\Psi_k)$, the semiclassical
measures $\rho_{stand}$
obtained as limits of their ``standard'' Husimi measures will differ from the Walsh semiclassical measure
$\rho_{Walsh}$ described in the theorems above, but their projections on the position axis will
be identical.
Notice that $\rho_{stand}$ has to be an eigenmeasure
of the multivalued baker's map, while $\rho_{Walsh}$ is an eigenmeasure of the (single-valued)
open baker's map. For a semiclassical sequence $(\Psi_k)$ with eigenvalues $|z|\to|z_{\max}|$
(resp. $|z|\to|z_{\min}|$),
it is not clear whether there is a unique semiclassical
measure $\rho_{stand}$, but in any case the projection $\pi_*\rho_{stand}$
is unique, equal to $\nu_{\Can}$ (resp. $\nu_{\min}$).

\medskip

Theorems 1-4 are our main results. The system we study is the first for which
these properties have been proved. It is natural to ask which, if any, of our
results extend to quantum chaotic scattering in general. At this stage, the
answer to this question is far from being clear, even heuristically, and we
consider it to be an interesting avenue for future investigations.

This paper is organized as follows. In the next section we describe the
Walsh-baker map in more detail. We prove the exact version of (\ref{weights}),
discuss the Walsh coherent states and the Walsh Husimi measures. In
section~\ref{s:psi-bet} we introduce the particular basis of eigenstates
$\{\Psi^m_{\bet}\}$ of the nontrivial spectrum and prove Theorems~\ref{th:2}
and \ref{th:3}, which involve the study of ``almost periodic'' binary sequences
(the properties that we need are derived in the Appendix). The eigenstates of
Theorem~\ref{th:4} are exhibited and studied in section~\ref{s:combination},
while the ``quantum uniqueness'' at the edges of the spectrum
(Theorem~\ref{th:1}) is proven in section~\ref{s:edge}.

\medskip

\noindent{\bf Acknowledgements} This work was supported by EPSRC. S.N. was
partially supported by the Agence Nationale de la Recherche, under the grant
ANR-05-JCJC-0107-01. He is grateful to Jens Marklof for organizing a session at
the School of Mathematics in Bristol, during which this collaboration was
initiated.

\section{Walsh quantization of the open baker map}

\subsection{Walsh kinematics and the open Walsh baker}\label{s:Walsh-kinematics}
In what follows we restrict ourselves to the triadic baker map.  A central
object in our analysis will be the Cantor set and its generations. Given the
interval $\Can_0=[0,1)$ the first such generation is
$\Can_1=[0,\frac{1}{3})\cup[\frac{2}{3},1)$, obtained by removing the middle
third $[\frac{1}{3},\frac{2}{3})$, which contains all points of the interval
such that their ternary decomposition starts with the digit `$1$'. Further
generations $\Can_n$ are obtained recursively, at each step removing the union
of the middle thirds of all the intervals. Therefore $\Can_n$ contains all
points for which the first $n$ digits in the ternary decomposition take value
in $\{0,2\}$. This process converges to a fractal of Hausdorff dimension $\ln
2/\ln 3$, known as the Cantor set, which we denote by $\Can=\Can_\infty$. This
set contains the points in $[0,1)$, the ternary decomposition of which is made
only of the symbols $\{0,2\}$.

As already mentioned, the allowed position values at the quantum
level are of the form $q_j=(j+1/2)/N$. We only consider the case
$N=3^k$, so that each of the $N$ positions can be labelled by $k$
symbols \be\label{ternary}
q_j-\frac1{2N}=0\cdot\epsilon_1\epsilon_2\cdots\epsilon_k\quad\text{if}\quad
j=\sum_{\ell=1}^{k}\epsilon_\ell\, 3^{k-\ell}, \qquad j\in
\{0,\ldots,N-1\},\quad \epsilon_\ell\in\{0,1,2\}\,. \ee The point
$q_j$ belongs to $\Can_k$ iff all the $\epsilon_\ell\neq 1$. This ternary
decomposition allows position eigenstates $|q_j\ra$ to be written
formally as tensor products, \be
|q_j\ra=|\epsilon_1\ra\otimes\cdots\otimes|\epsilon_k\ra,\qquad
\{|\eps\ra,\ \eps=0,1,2\}\quad\text{being the standard basis in } \IC^3\,. \ee
This expression takes advantage of the
particular structure of the triadic baker map. It is very convenient
for addressing the generations of the Cantor set, and constructing
the eigenstates of $B$. The kernel of the Walsh-Fourier transform in
the position basis is, for any $j,j'=0,\ldots,N-1$, \be\label{walsh}
(W_N)_{jj'}=3^{-k/2}\prod_{\ell=1}^k\exp\left(-\frac{2i\pi}{3}
(\epsilon_\ell(j)+1/2)(\epsilon_{k+1-\ell}(j')+1/2)\right)\,. \ee
Equivalently, the action of this transform on tensor product states
is given by \cite{jpa38sn2005,cmp269sn2007} \be
W_N(v_1\otimes\cdots\otimes v_k)=F_3v_k\otimes\cdots\otimes F_3v_1,
\ee where $v_i\in\mathbb{C}^3$ stand for any linear superposition of
the basis $\{|\eps\ra,\ \eps=0,1,2\}$, and $F_3$ is the
$3\times 3$ matrix defined in \eqref{fourier}. Mimicking the
``standard'' quantization \eqref{quantumbaker} of the open triadic
baker, one gets the matrix \eqref{WalshB} as the Walsh-quantization
of that map. A similar quantization for the closed baker map
originally appeared in the context of quantum information
\cite{schack}. More recently, a wide variety of quantizations of the
baker map have been systematically studied \cite{pre74el2006}.

The matrices $B$, $B^\dag$ preserve the tensor product
decomposition, acting as twisted shifts:
\begin{eqnarray}\label{Baction}
B(v_1\otimes\cdots\otimes
v_k)&=&v_2\otimes\cdots\otimes v_k\otimes\widetilde{F}_3^\dag v_1\,,\\
B^\dag(v_1\otimes\cdots\otimes v_k)&=&\widetilde{F}_3 v_k
\otimes v_1\otimes\cdots\otimes v_{k-1}\,.
\end{eqnarray}
Here $\widetilde{F}_3$ is the ``open" Fourier transform, obtained by setting to
zero the middle row of $F_3$.

\subsection{Spectrum of the open Walsh baker}\label{s:spectrum}

The action of the $k$-th iterate $B^k$ is \be \label{tensor}
B^k(v_1\otimes\cdots\otimes v_k)=\tF_3^\dag v_1\otimes\cdots\otimes \tF_3^\dag
v_k, \ee so we first need to diagonalize the $3\times 3$ matrix $\tF_3^\dag$.
We call the right eigenvectors of $\tF_3^\dag$ \be\label{frs}
|f_-\ra=\frac{1}{\sqrt{2}}(1,0,-1)^T, \quad
|f_+\ra=\frac{1}{\sqrt{6}}(1,2,1)^T, \quad |f_0\ra=(0,1,0)^T, \ee the
subscripts being inspired by the last entry of the vector. An important
property is the {\em orthogonality} of $|f_-\ra$ and $|f_+\ra$, which is
specific to the choice of antiperiodic boundary conditions (the choice of
periodic boundary conditions yields nonorthogonal vectors \cite{Rubin}). The
respective eigenvalues are \be\label{values} \lambda_-=1, \quad
\lambda_+=\lambda=\frac{i}{\sqrt{3}}, \quad \lambda_0=0. \ee The left
eigenvectors of $\tF_3^\dag$ are \be\label{gs} \la
g_-|=\frac{1}{\sqrt{2}}(1,0,-1), \quad \la g_+|=\frac{i}{\sqrt{2}}(1,0,1),
\quad \la g_0|=\frac{i}{\sqrt{3}}(1,-1,1). \ee Left and right eigenstates are
related by \be\label{ltor} F_3|f_0\ra=|g_0\ra, \quad F_3|f_-\ra=|g_-\ra, \quad
F_3|f_+\ra=|g_+\ra. \ee In particular, we see that the position representation
of left eigenstates is equivalent to the momentum representation of right
eigenstates. We may thus restrict our attention to right eigenstates only.

The eigenvalues of $B^k$ are obviously given by products of the eigenvalues in
(\ref{values}), and we can make the following sharp distinction between
short-lived and long-lived eigenstates:

{\bf Convention:} {\it Eigenstates of $B$ corresponding to non-zero
eigenvalues are called long-lived, and the remaining ones
short-lived.}

The nontrivial spectrum of $B^k$ is spanned by a subspace of dimension $2^k$.
The $k+1$ eigenvalues $\{\lambda^d\,:\,d=0,\ldots,k\}$ are in general highly
degenerate: the multiplicity of $\lambda^d$ is the binomial coefficient
$\binom{k}{d}$. The nonzero eigenvalues of $B$ are then simply of the form
$\lambda^{d/k}e^{2\pi im/k}$ with $0\leq m\leq k-1$. The largest one in modulus
is $z_{max}=1$, while the smallest is $z_{min}=\lambda=i/\sqrt{3}$. The
remaining ones form a lattice inside the annulus $\{|z_{min}|=1/\sqrt{3}\leq
|z|\leq z_{max}=1\}$, which becomes circular symmetric in the limit
$k\to\infty$. No Jordan block appears in the nontrivial spectrum of $B$.

As noticed in \cite{jpa38sn2005,cmp269sn2007}, the number of nonzero
eigenvalues of $B$ (counted with multiplicities) scales like $N^{\ln
2/\ln 3}$, which corresponds to the fractal Weyl law for this system
($\ln 2/\ln 3$ is half the dimension of the repeller $K_0$).

\subsection{Exact scaling properties of the eigenstates}

Let us define the $3\times 3$ ``in" and ``out" projectors
\be
\pi_I=\begin{pmatrix}
                  1 &   &   \\
                    & 0 &   \\
                    &   & 1
\end{pmatrix}, \quad
\pi_O=1-\pi_I=\begin{pmatrix}
                  0 &   &   \\
                    & 1 &   \\
                    &   & 0
           \end{pmatrix} ,
\ee  and the tensor products \be
\Pi_n=\underbrace{\pi_I\otimes\pi_I\otimes\cdots\otimes\pi_I}_{n-1}\otimes\pi_O
\otimes\underbrace{1\otimes\cdots\otimes1}_{k-n}, \quad 1\leq n\leq k. \ee We
see that $\Pi_1$ is the projector onto the hole. Let us also define the set of
points which fall into the hole after $n$ steps, but not earlier, \be
\mathcal{R}_n=\{x\in {O}_{-n},\,x\notin {O}_{-m}, \,0\leq m<n\},
\ee with the convention that $\mathcal{R}_0={O}$. These sets can be
written as $\mathcal{R}_n=(\Can_n\setminus\Can_{n+1})\times[0,1)$, and are
related by \be\label{rs} \mathcal{U}^{-1}(\mathcal{R}_n)\setminus
{O}=\mathcal{R}_{n+1}. \ee With this definition the projector $\Pi_n$
corresponds to the region $\mathcal{R}_{n-1}$.

The semiclassical propagation of these projectors is exact up to the Ehrenfest
time $k$, in the sense that $B^\dag\Pi_n B=\Pi_{n+1}$ for any $n<k$. Indeed,
the left hand side acts as follows on a tensor product state ${\bf v}$: \be
B^\dag\,\Pi_n\,B\,{\bf v}=
(\pi_Iv_1)\otimes(\pi_Iv_2)\cdots(\pi_Iv_{n})\otimes(\pi_Ov_{n+1})\otimes\cdots
\otimes v_k=\Pi_{n+1}{\bf v}. \ee For any $n\leq k$ we thus have
$\Pi_n=(B^\dag)^{n-1}\Pi_1B^{n-1}$, and the fact that $\Pi_1=1-B^\dag B$ leads
to the relation $\Pi_n=(B^\dag)^{n-1}B^{n-1}-(B^\dag)^{n}B^{n}$. Let $|\Psi\ra$
be a normalized right eigenstate of $B$ with eigenvalue $z$. This state then
satisfies \be \forall n,\  0\leq n<k,\qquad
\la\Psi|\Pi_{n+1}|\Psi\ra=|z|^{2n}(1-|z|^{2}), \ee which is the exact version
of the general semiclassical property \eqref{weights}.

\subsection{Walsh Coherent States}\label{s:WalshCS}
In order to investigate the phase space distribution of the eigenstates of the
Walsh-quantized open baker map, we use Walsh-coherent states
\cite{thiele,entropy} and the associated Walsh-Husimi representations of
quantum states. While the usual coherent states are associated with Gaussians
in phase space, the `Walsh' ones are associated with ``quantum rectangles''.

We have been denoting position eigenstates by
$|q_j\ra=|\epsilon_1\ra\otimes\cdots\otimes|\epsilon_k\ra$.
The action of the Walsh-Fourier transform on these states yields the
orthonormal basis of Walsh-momentum eigenstates. Given a certain momentum
$p_j=(j+1/2)/N$, with $j=\epsilon'_1\epsilon'_2\cdots\epsilon'_k$ in
ternary notation, we associate to $p_j$ the state
\be
|p_j\ra \defeq W_N^\dag|q_j\ra=F_3^\dag|\epsilon'_k\ra\otimes
F_3^\dag|\epsilon'_{k-1} \ra\otimes\cdots\otimes
F_3^\dag|\epsilon'_1\ra.
\ee
As explained in \S\ref{s:Walsh-baker}, given an integer $0\leq b\leq k$,
two sequences $\bep=\eps_1\cdots\eps_b$,
$\bep'=\eps'_1\cdots\eps'_{k-b}$ define a ``quantum rectangle'' $[\bep'\cdot\bep]$
of size $b\times (k-b)$.
To this rectangle we associate the
Walsh-coherent state $|\bep'\cdot\bep\ra_b$:
\be
|\bep'\cdot\bep\ra_b\defeq |\eps_1\ra\otimes\cdots\otimes|\eps_b\ra\otimes
F_3^\dag|\eps'_{k-b}\ra\otimes \cdots\otimes
F_3^\dag|\eps'_1\ra.
\ee
In particular, for $b=0$ we recover
momentum eigenstates, while $b=k$ corresponds to position
eigenstates.

For each choice of $0\leq b\leq k$, the family of coherent states
$\{|\bep'\cdot\bep\ra_b\}$ forms an orthonormal basis of the
quantum Hilbert space.
Once we select the parameter $b$,
the {\em Walsh-Husimi measure} associated with a normalized state
$|\psi\ra$ is a probability measure
defined through its values on the rectangles of size $b\times (k-b)$:
\be
H^b_\psi([\bep'\cdot\bep])=
|\la\psi|\bep'\cdot\bep\ra_b|^2\,. \ee In the semiclassical limit a
sequence of quantum rectangles can converge to a phase space point
only if the parameter $b$ depends on $k$ in such a way that
$b(k)\to\infty$ and $k-b(k)\to\infty$. The ``most isotropic'' choice
consists in taking $b=[k/2]$. Under these conditions, any sequence
$(H^{b(k)}_{\Psi(k)})_{k\to\infty}$ admits one or several weak-$*$
limit measures $\rho$ on the torus, and $\rho$ does not depend on
the precise choice of $b$ \cite{entropy}.

To study the semiclassical limits, we need to compute the weights of
$H^{b(k)}_{\Psi(k)}$ on fixed (that is, $k$-independent) rectangles, for
instance on the family of $\vis$-squares $[\bep'\cdot\bep]_{\vis}$ for some
fixed $\vis\in\IN$ (we will sometimes call these squares ``classical'' to
insist on the independence with respect to $k$). Let
$\Pi_{[\bep'\cdot\bep]_{\vis }}$ denote the projector \be\label{e:proj-v}
\Pi_{[\bep'\cdot\bep]_{\vis
}}=\pi_{\eps_1}\otimes\cdots\pi_{\eps_{\vis}}\otimes I\cdots
I\otimes\tilde\pi_{\eps'_{\vis}}\otimes\cdots\otimes\tilde\pi_{\eps'_1} \ee
(where $\pi_{\eps}=|\eps\ra\la\eps|$, $\tilde\pi_{\eps}=F_3^\dagger \pi_\eps
F_3$) associated with the square $[\bep'\cdot\bep]_{\vis }$. For any parameter
$b\ge\vis$ such that $k-b\ge v$, the value of the Husimi measure $H^b_{\psi}$
on $[\bep'\cdot\bep]_{\vis }$ is defined to be \be\label{Honvquare}
H_{\psi}^b([\bep'\cdot\bep]_{\vis }) = \la\psi|\Pi_{[\bep'\cdot\bep]_{\vis
}}|\psi\ra\,. \ee Notice that this is actually independent of $b$.
\medskip

Using the fact that $F_3^2={\scriptsize
\begin{pmatrix}&&-1\\&-1&\\-1&&\end{pmatrix}}$, the Walsh-Fourier transform
(\ref{walsh}) acts as follows on a coherent state: \be
W_N|\bep'\cdot\bep\ra_b=|\eps'_1\ra\otimes \cdots\otimes|\eps'_{k-b}\ra\otimes
F_3|\eps_b\ra\otimes\cdots\otimes F_3|\eps_1\ra=
(-1)^b\,|\bar\bep\cdot\bep'\ra_{k-b}\,, \ee where we defined
$\bar\eps_j=2-\eps_j$ for all $j=1,\ldots,b$. The interval indexed by the
sequence $\bar\bep$ is symmetrical (with respect to the origin) to the one
indexed by $\bep$. Thus, the rectangle $[\bar\bep\cdot\bep']$ is the image of
the rectangle $[\bep'\cdot\bep]$ after a phase space rotation of $\pi/2$ around
the origin. We have recovered the Walsh analogue of the action of the Fourier
transform $F_N$ on Gaussian coherent states.

The operator $B$ maps $b$-coherent states into $(b-1)$-coherent states. Indeed,
from (\ref{Baction}) we have (if $b\geq 1$) \be
B|(\eps'_{k-b}\cdots\eps'_{1})\cdot
(\eps_1\cdots\eps_b)\ra_b=\delta_{\eps_1\neq 1}\;
|(\eps'_{k-b}\cdots\eps'_{1}\eps_1)\cdot(\eps_2\cdots\eps_b)\ra_{b-1}. \ee This
action exactly corresponds to the action of the classical open baker map $\tcU$
on the rectangles $[\bep'\cdot\bep]$: a rectangle inside the hole ($\eps_1=1$)
is ``killed'', while a rectangle outside the hole is transformed classically:
\begin{align*}
B|\bep'\cdot\bep\ra_b&=\delta_{\eps_1\neq 1}\;|\cU([\bep'\cdot\bep])\ra_{b-1},
\quad\text{and similarly}\\
B^\dag|\bep'\cdot\bep\ra_b&=
\delta_{\eps'_1\neq 1}\;|\cU^{-1}([\bep'\cdot\bep])\ra_{b+1}\,.
\end{align*}
Let $|\Psi\ra$ be a long-lived right eigenstate, with $B|\Psi\ra=z|\Psi\ra$,
$z\neq 0$. Then for any rectangle such that all $\eps'_j\neq 1$, we have
\be\label{e:invariance} \forall n\leq k-b,\qquad
H^b_{\Psi}([\bep'\cdot\bep])=|z|^{-2n}\;H_\Psi^{b+n}(\cU^{-n}
([\bep'\cdot\bep]))\,. \ee If on the contrary one of the symbols $\eps'_j=1$,
then $H^b_{\Psi}([\bep'\cdot\bep])=0$. This is the case iff the rectangle
$[\bep'\cdot\bep]$ escapes the system when propagated backwards. The Husimi
measure $H^b_{\Psi}$ is thus supported on $[0,1)\times\Can_{k-b}$, which can be
seen as a coarse-grained version of the backward-trapped set $K_+$. The
concentration of the Husimi function on $K_+$ was discussed in
\cite{casati,previous,Rubin} for the semiclassical limit. The arguments above
show that for this system a precise localization already holds for finite $k$.

The covariance \eqref{e:invariance} implies the following (Egorov-type)
estimate on the weights of $\vis$-squares. Assume $k\ge4\vis$ and take an
arbitrary eigenstate $\Psi$ with eigenvalue $z\neq 0$. Its Husimi measure
satisfies: \be\label{e:Egorov1} H^{[k/2]}_{\Psi}([\bep'\cdot\bep]_{\vis })  =
|z|^{-2\vis}\, \big(\prod_{i=1}^{\vis}\delta_{\eps'_i\neq1}\big)\,
H^{[k/2]}_{\Psi}([\cdot\bep'\bep]_{2\vis })\,. \ee Hence, to compute the
weights of $\vis$-squares one only needs to know the weights of ``vertical''
rectangles $[\cdot\bep'\bep]$ of size  $2\vis\times 0$. For actual
computations, it will prove convenient to use this property.

\section{A particular family of long-lived eigenstates}\label{s:psi-bet}

Let us first describe the short-lived (right) eigenstates of $B$.
From the action of $B$ on tensor products \eqref{Baction}, we see
that if ${\bf v}=v_1\otimes\cdots\otimes v_k$ with $v_1=|f_0\ra$
then $B{\bf v}=0$. There are $N/3$ such degenerate states, and by
taking their overlaps with coherent states we see that they are
supported in the hole $\mathcal{R}_0$. If $v_1\neq|f_0\ra$ but
$v_2=|f_0\ra$ then ${\bf v}$ is not annihilated by the action of
$B$, but we now have $B^2{\bf v}=0$, so ${\bf v}$ is an eigenstate
of $B$ in a generalized sense. Since $|f_0\ra$ appears in the second
position, the (Walsh-)Husimi function $H^b_{\bf v}$ of this state is
localized in the set $\mathcal{R}_1$. More generally, the Husimi
function of a state ${\bf v}$ such that $B^n{\bf v}=0$ and $B^m{\bf
v}\neq 0$ for $0<m<n$ is supported on the set $\mathcal{R}_{n-1}$
(provided this Husimi function is defined with a parameter $b\geq
n$).

Hence the short-lived states are supported on classical phase space
regions that escape the system before $k$ steps ($k$ corresponds
to the Ehrenfest time). In the present system this escape is
perfectly deterministic and thus the short-lived states span the
generalized kernel of $B$. For more general systems some unavoidable
leakage will lift this degeneracy and lead to small but finite
eigenvalues (see \cite{henning}).

In the rest of the paper we will focus on the long-lived
eigenstates. Since the corresponding eigenspaces are generally quite
degenerate, we will choose a particular eigenbasis, which we now
describe.

\subsection{Construction of the eigenstates $\Psi_{\bet}^m$}\label{s:eigenstates}

As explained in section~\ref{s:statement}, we denote by
$\bet=\eta_1\eta_2\cdots\eta_k$, with $\eta_j\in\{+,-\}$, a binary sequence of
length $k$. The number of times the positive sign appears in the sequence
$\bet$ is called its {\em degree} $d$. The ratio  $\delta\defeq d/k$ will be
called the {\em relative degree} of $\bet$. To each sequence $\bet$ we
associate the state \be\label{eta} |\bet\ra=|f_{\eta_1}\ra\otimes
|f_{\eta_2}\ra\otimes\cdots\otimes |f_{\eta_k}\ra, \ee where $|f_{\pm}\ra$ are
the eigenvectors of $\tF_3^\dag$ given in \eqref{frs}. The state $|\bet\ra$ is
a right eigenstate of $B^k$, with the eigenvalue $\lambda^{d}$.

Let $\tau$ denote the cyclic shift, such that
$\tau\bet=\eta_2\cdots\eta_k\eta_1$. Each sequence has a minimal period
$1\leq\ell\leq k$ under $\tau$ such that $\tau^\ell\bet=\bet$. This period
obviously divides $k$. The orbit of $\bet$ is the set $[\bet]=\{\tau^j\bet,\
0\leq j\leq\ell-1\}$ (if $\bet,\bet'$ belong to the same orbit, we will write
$\bet\equiv\bet'$). For a given pair $\ell,d$ there may exist more than one
orbit. For example, for $k=5,d=2$ the sequences ``$++---$" and ``$+-+--$"
belong to different orbits, both of period $\ell=5$.

The action of $B$ on $|\bet\ra$ can be written as
$B|\bet\ra=\lambda_{\eta_1}|\tau\bet\ra$. Each orbit $[\bet]$ provides us with
a family of eigenstates of $B$. Let us select one representative $\bet$ in this
orbit. For any $0\leq m\leq\ell-1$ we define
$z_{\delta,m/\ell}=\lambda^{\delta}\,e^{2\pi im/\ell}$ and construct the state
\be \label{e:Psi-dm} |\Psi^{m}_{\bet}\ra=
\frac{1}{\sqrt{\cN}}\sum_{j=0}^{\ell-1}\frac{B^j}{z_{\delta,m/\ell}^j}|\bet\ra
=\frac{1}{\sqrt{\cN}}\sum_{j=0}^{\ell-1}c_{jm} |\tau^j\bet\ra,\ee
where\be\label{coeffs}
c_{jm}=\prod_{s=1}^{j}\frac{\lambda_{\eta_s}}{z_{\delta,m/\ell}}, \quad
c_{0m}=1,\ee and \be
\cN=\sum_{j=0}^{\ell-1}|c_{jm}|^{2}.\label{e:normalization} \ee This state is a
(right) eigenstate of $B$ with eigenvalue $z_{\delta,m/\ell}$. This can be
verified by direct inspection. Up to a global phase, it only depends on the
orbit $[\bet]$, so with some abuse we may call it $\Psi^{m}_{[\bet]}$. Notice
that $|c_{jm}|$ is independent of $m$, and so is $\cN$. Due to the
orthogonality between $|f_-\ra$ and $|f_+\ra$, the state $\Psi^{m}_{[\bet]}$ is
normalized, and two states $\Psi^{m}_{[\bet]}$, $\Psi^{m'}_{[\bet]}$ with
$m\neq m'$ are orthogonal to each other. In the same way, eigenstates
constructed from different orbits $[\bet]\neq [\bet']$ are also orthogonal to
each other. The family
$$
\{\Psi^{m}_{[\bet]},\,:\, [\bet],\, m=0,\ldots,\ell(\bet)-1 \}
$$
thus forms an orthonormal basis of the nontrivial spectrum of $B$.

If $m,\ell$ are coprime, the degeneracy of the eigenvalue
$z_{\delta,m/\ell}$ is the number of different orbits with length
$k$, degree $d$ and periods $\ell'$ such that $\ell|\ell'$ and
$\ell'|k$.

\subsection{Spectral averages of Husimi measures}\label{s:4}

In this section we prove Theorem~\ref{th:2}, which describes spectral averages
of Husimi functions of the form \eqref{e:average-Hus}. Using the eigenfunctions
we have constructed, these averages take the form \be
H^b_{[\bet]}([\bep'\cdot\bep])=\frac{1}{\mathcal{N}\ell}\sum_{m=0}^{\ell-1}\left|\sum_{j=0}^{\ell-1}\overline{c_{jm}}\langle
\tau^j\bet |\bep'\cdot\bep\rangle_b\right|^2.\ee

For any $0\leq j,j'< \ell$, one has \be\label{media}
\frac{1}{\ell}\sum_{m=0}^{\ell-1}(z_{\delta,m/\ell}^\ast)^{-j}\,(z_{\delta,m/\ell})^{-j'}=
\delta_{j,j'}\,|\lambda|^{- 2j\delta}\,, \ee so that
$\sum_{m=0}^{\ell-1}\overline{c_{jm}}\,c_{j' m}=0$ if $j\neq j'$. Averaging
over $m$ thus cancels off-diagonal terms. For any quantum rectangle
$[\bep'\cdot\bep]$ of size $b\times(k-b)$ we find \be\label{average2}
H^b_{[\bet]}([\bep'\cdot\bep])= \cN^{-1}\sum_{j=0}^{\ell-1}|c_{j0}|^{2}\,|\la
\tau^j\bet |\bep'\cdot\bep\ra_b|^2 \,. \ee Each overlap on the right hand side
takes the value \be
 | \la \tau^j\bet | \bep'\cdot\bep\ra_b |^2 =
\prod_{i=1}^{k-b}|\la\eps'_i|g_{\eta_{j-i+1}}\ra|^2\,
\prod_{i=1}^{b}|\la\eps_i|f_{\eta_{i+j}}\ra|^2 \ee (the indices $\eta_{i+j}$
are extended by periodicity). Let us first study the dependence of
$H_{[\bet]}^b([\bep'\cdot\bep])$ with respect to the momentum coordinate (that
is, the symbols $\eps'_i$). From the expressions \eqref{gs} for $|g_{\pm}\ra$,
we immediately see that $|\la\eps'_i|g_{\eta_{j-i+1}}\ra|^2=1/2$ if
$\eps'_i\neq 1$, and that it vanishes otherwise, independently of $\bet$ or
$j$. The momentum dependence can thus be factorized:
\begin{align}
H^b_{[\bet]}([\bep'\cdot\bep])&=\nu_{\Can}([\bep'])\;\nu_{[\bet]}([\bep]),\qquad\text{where}
\label{e:factorization}\\
\nu_{[\bet]}([\bep])&=\cN^{-1}\sum_{j=0}^{\ell-1}|c_{j0}|^{2}\,
\prod_{i=1}^{b}|\la\eps_i|f_{\eta_{i+j}}\ra|^2\,.\label{e:factor}
\end{align}
Here $\nu_{\Can}$ is the ``uniform measure'' on $\Can$ defined in
\eqref{e:nu_Can}. If we extend formula \eqref{e:factor} to sequences $\bep$ of
arbitrary length, it specifies a probability measure $\nu_{[\bet]}$ on the unit
interval, and therefore also a probability measure
$\rho_{[\bet]}=\nu_{\Can}(dp)\,\nu_{[\bet]}(dq)$ on the torus. The averaged
Husimi measure $H^b_{[\bet]}$ is equal to $\rho_{[\bet]}$, conditioned on the
rectangles $[\bep'\cdot\bep]$ of type $b\times (k- b)$. This proves the first
statement of Theorem~\ref{th:2}.

From \eqref{average2}, we see that $\nu_{[\bet]}$ is a convex combination of
measures $\nu_{\tau^j\bet}$, where \be\label{e:nu-tau}
\nu_{\tau^j\bet}([\eps_1\cdots\eps_b])=\prod_{i=1}^{b}|\la\eps_i|f_{\eta_{i+j}}\ra|^2
\ee is a Bernoulli measure associated with the sequence $\tau^j\bet$.
$\rho_{[\bet]}$ thus belongs to the class of eigenmeasures studied in
\cite[Prop.8]{Rubin}; in particular, it is conditionally invariant through the
open map $\tcU$: \be\label{e:invar}
\tcU^{*}\,\rho_{[\bet]}=|\lambda|^{2\delta}\,\rho_{[\bet]}\,. \ee Inserting
\eqref{e:invar} in the decomposition \eqref{e:factorization}, we obtain the
scaling relation \eqref{e:scaling}. Sequences $\bep\in\{0,2\}^n$  correspond to
intervals of $\Can_n$. The fact that \eqref{e:factor} is independent of the
choice of the subsequence $\bep\in\{0,2\}^n$ shows that the measure
$\nu_{[\bet]}$ has the same shape in each connected component of $\Can_n$.
From the expressions \eqref{frs} we see that for any $b$-sequence $\bep$ such
that exactly $n$ symbols satisfy $\eps_i\neq 1$, one has
$\nu_{\tau^j\bet}([\bep])\leq \frac1{2^{n}}\,\big(\frac23\big)^{b-n}$. The same
inequality obviously applies to the convex combination $\nu_{[\bet]}$, which
proves \eqref{e:bounds}. This ends the proof of Theorem $2$.

Through the scaling property \eqref{e:scaling}, we see that the measure
$\nu_{[\bet]}$ can be specified by its shape inside the hole (this is a general
property of conditionally invariant measures with $e^{-\Gamma}<1$
\cite{n19mfd2006}). This shape depends on the specific orbit $[\bet]$. For
instance, Figure~\ref{f:fig3} shows the weights $\nu_{[\bet]}([\bep])$ for
sequences $\bep$ of length $5$. In that figure, another obvious property of
$\nu_{[\bet]}$ is its symmetry with respect to the middle point $q=1/2$. This
property is easy to check in terms of symbolic sequences. For any sequence
$\bep$, let $\bar\bep$ be the sequence obtained from $\bep$ by replacing
everywhere $0$ by $2$ and vice-versa. The interval $[\bar\bep]$ is exactly the
symmetric partner of $[\bep]$ with respect to the middle point. Then, one
easily checks that for any sequence $\bep$,
$\nu_{[\bet]}([\bar\bep])=\nu_{[\bet]}([\bep])$.

\subsection{Husimi weights of ``classical rectangles''}\label{s:husimi}

In the following sections we will compute Husimi weights of classical rectangles.
Keeping $\vis>0 $ fixed, we select for each $k\ge2\vis$ an eigenstate
$\Psi^{m}_{\bet}$ of the form \eqref{e:Psi-dm}. For convenience, we will
consider the ``isotropic'' Husimi measures $H_{\Psi}=H^{[k/2]}_{\Psi}$. As
explained before, the sequence $(H_{\Psi^{m}_{\bet}})_{k\to\infty}$ has a
chance to converge to a semiclassical measure only if the sequences
$\bet=\bet(k)$ are chosen such that their relative degrees $\delta(k)\to
\delta$.

If the periods $\ell=\ell(k)$ of the sequences $\bet(k)$ are uniformly bounded,
we may use the results of \cite{Rubin} to classify the semiclassical measures.
Indeed, if $\tbet$ is a fixed, primitive sequence, then the Husimi measures
associated to the states $\Psi^m_{(\tbet)^{k'}}$ (with $k'\to\infty$ and
$m=m(k')$ arbitrary) converge to the measure $\nu_{\Can}(dp)\times
\nu_{[\tbet]}(dq)$ described in \eqref{e:factorization} and below.
We will thus concentrate here on sequences $\bet(k)$ of periods $\ell(k)\to\infty$.

From the invariance property \eqref{e:Egorov1}, we may restrict our
investigation to the weights of vertical rectangles
$[\cdot\bep]_{\vis}=[\cdot\eps_1\cdots\eps_{\vis}]$. For any
primitive $k$-sequence $\bet$ and any $m\in\{0,\ldots,k-1\}$, the
Husimi measure $H_{\Psi^{m}_{\bet}}$ of such a vertical rectangle
reads \be\label{e:coarse} H_{\Psi^{m}_{\bet}}([\cdot\bep]_{\vis })
=\cN^{-1}\sum_{j,j'=0}^{k-1} \overline{c_{jm}}\,c_{j'm} \la
\tau^{j}\bet|\Pi_{[\cdot\bep]_{\vis }}|\tau^{j'}\bet\ra. \ee Each
term $(j,j')$ in the right hand side of \eqref{e:coarse} contains
the factor \be \prod_{i=\vis +1}^{k} \la
f_{\eta_{i+j}}|f_{\eta_{i+j'}}\ra\,. \ee From the
orthogonality $\la f_-|f_+\ra=0$, this factor vanishes unless the
sequences $\tau^{j}\bet$ and $\tau^{j'}\bet$ coincide along the
index set $\{\vis+1,\ldots,k\}$ (the ``$\vis$-bulk''), or
equivalently, outside the set $\{1,\ldots,\vis\}$ (the
``$\vis$-box'').

If $\bet$ is not primitive, that is if $\bet=\tbet^n$ for some primitive
$\tbet$ and $n>1$, then as soon as $k\ge2\vis$ the $v$-bulk of $\tau^j\bet$ will
always contain a {\it full} sequence $\tau^j\tbet$: this implies that the terms
$\la\tau^j\bet|\Pi_{[\cdot\bep]_{\vis }}|\tau^{j'}\bet\ra$ vanish if $j\neq
j'\bmod\ell$, and one has
$H_{\Psi^{m}_{\bet}}([\cdot\bep]_{\vis })=H_{[\bet]}([\cdot\bep]_{\vis })$.
For this reason we will from now on restrict our attention
to eigenstates constructed from long {\em primitive} sequences $\bet$.

\noindent{\bf Definition.} For $\vis>0$ fixed, we take $k\geq\vis$
and consider primitive sequences $\bet=\eta_1\cdots\eta_k$.  If
there exist two {\it different} integers $j,j'\in\{0,\ldots,k-1\}$
such that the sequences $\tau^j\bet$ and $\tau^{j'}\bet$ coincide on
the $\vis$-bulk, then the sequence $\bet$ is said to be
($\vis$-){\it admissible}. The pair $(j,j')$ is then called an {\it
admissible pair} for $\bet$, and we write
$j\stackrel{\vis,\bet}{\sim}j'$.
Obviously, admissibility is a property of the orbit $[\bet]$.

\medskip

The Husimi weight \eqref{e:coarse} can be decomposed into:
\be\label{e:decompo0} H_{\Psi^{m}_{\bet}}([\cdot\bep]_{\vis })
=\cN^{-1}\sum_{j=0}^{k-1} |c_{jm}|^2 \la \tau^{j}\bet|\Pi_{[\cdot\bep]_{\vis
}}|\tau^{j}\bet\ra + \cN^{-1}\sum_{j\stackrel{\vis,\bet}{\sim}j'}
\overline{c_{jm}}\,c_{j' m}\, \la \tau^{j}\bet|\Pi_{[\cdot\bep]_{\vis
}}|\tau^{j'}\bet\ra\,. \ee This weight is thus made of ``diagonal'' and
``off-diagonal'' terms. We have analyzed the former in the previous subsection.
Our main task will now consist in estimating the contribution of the latter in
the cases where it is nontrivial (that is, when $\bet$ is $\vis$-admissible).

\subsection{Semiclassical measures of the individual eigenstates $\Psi^{m}_{\bet}$}

Unlike in the last section, we now fix $\vis>0$ and focus on the
{\it individual} Husimi weights
$H_{\Psi^{m}_{\bet}}([\cdot\bep]_{\vis })$ given in
\eqref{e:decompo0}, in the limit $k\to\infty$. The previous section
described some properties of the diagonal sum in \eqref{e:decompo0}.
In this section (which strongly depends on the Appendix), we show
that the off-diagonal sum in \eqref{e:decompo0} is always negligible
in the semiclassical limit, as long as one considers the weights of
``classical'' rectangles (that is, take $\vis$ fixed and
$k\to\infty$).
\begin{prop} \label{p:individual}
Fix $\vis \geq 1$ and take any vertical $\vis$-square $[\cdot\bep]_{\vis}$.
Then there exists a constant $C_{\vis}$ such that the following holds. For any
$k\ge2\vis$, take a primitive $k$-sequence $\bet$ and an arbitrary $m\in
\{0,\ldots,k-1\}$. One has then: \be\label{e:off-diag} \Big|\cN(\bet)^{-1}
\sum_{j\stackrel{\vis,\bet}{\sim}j'}\overline{c_{jm}}\,c_{j' m}\, \la
\tau^{j}\bet|\Pi_{[\cdot\bep]_{\vis }}|\tau^{j'}\bet\ra\Big|\leq
C_{\vis}\,k^{-1}\,. \ee
\end{prop}
Before proving this proposition, we briefly explain how it yields
Theorem~\ref{th:3}. If the sequence $\bet$ appearing in Thm.~\ref{th:3} is
primitive, this is a bound on the off-diagonal sum in \eqref{e:decompo0}. As
already discussed, if $\bet$ is not primitive and $k\ge2\vis$ the off-diagonal
terms vanish. Besides, the arguments in section \ref{s:4} show that the
diagonal terms in \eqref{e:decompo0} yield $\nu_{[\bet]}([\cdot\bep]_{\vis})$,
so we get \eqref{e:stat-indiv} in the case of vertical $\vis$-rectangles.
Finally, from the Egorov property \eqref{e:Egorov1} the same equation holds if
we replace a vertical $2\vis$-rectangle by a $\vis$-square
$[\bep'\cdot\bep]_{\vis}$.

{\it Proof of the Proposition.} The results of \S\ref{s:husimi} show
that the right hand side in \eqref{e:off-diag} can be nonvanishing
only if the $k$-sequence $\bet$ is $\vis$-admissible. Among the full
set of primitive sequences of length $k$, admissible sequences
constitute a very restricted set: even though they are
primitive, these sequences are {\it almost periodic}, and
have a rich hierarchical structure, described in the Appendix.
We now describe some features of this almost-periodic structure, relevant for our aims.

\subsubsection{Hierarchical structure of admissible sequences}\label{s:hierarchy}
Fix $\vis>0$ and $k\gg\vis$. The analysis of the Appendix classifies
the family of $\vis$-admissible binary sequences (which are
primitive of length $k$) according to their {\it rank}, which is a
positive integer $n\leq \log_2 k$. The rank describes the
number of {\it levels} used to encode the hierarchical structure of
the sequence.

A $\vis$-admissible sequence of rank $n$ will be a repetition of two
``elementary strings'', which we will denote by $R_n$ and $D_n$. The
letters $R,\,D$ are for ``Repeated'' vs. ``Defect'', while the
subscript $n$ means that these strings correspond to the
``$n$-th level'' of $\bet$. The strings $D_n$, $R_n$ have lengths
$\leq\vis$, and they cannot be of the form $R_n=(\tbet)^m$,
$D_n=(\tbet)^{m'}$, that is repetitions of a common string $\tbet$.
We believe that the strings $D_n$, $R_n$ satisfy further
constraints, but we do not need to know these explicitly for our
purposes.

To construct the full sequence $\bet$ starting from the two strings $R_n$,
$D_n$, one proceeds iteratively from level $n$ down to level $1$. The
construction is encoded by a sequence of $n$ signed integers
\be\label{e:skeleton} (\sigma_1 r_1,\sigma_2 r_2,\ldots,
r_n),\quad\text{with}\quad \ \  r_i\geq 2,\ \  \sigma_i\in\{\pm\}. \ee Starting
from $j=n$ down to $j=1$, we use the two level-$j$ strings $D_j,\,R_j$ to
construct the ``long'' and ``short'' strings at level $j-1$ by the following concatenations:
\be\label{e:iteration1} \binom{L_{j-1}}{S_{j-1}}\defeq
\binom{D_j\,R_j^{r_j}}{D_j\,R_j^{r_j-1}}\,. \ee One of these two level-($j-1$)
strings will be the ``defect'', the other one being the ``repeated string'';
the choice depends on the sign $\sigma_{j}$: \be\label{e:iteration2} \forall
j=2,\ldots,n,\qquad
\begin{cases}D_{j-1} = L_{j-1},\ \ R_{j-1} = S_{j-1} &\text{ if }\sigma_{j-1}=+\,,\\
D_{j-1} = S_{j-1},\ \ R_{j-1} = L_{j-1} &\text{ if }\sigma_{j-1}=-\,.
\end{cases}
\ee
Finally, the $k$-sequence $\bet$ is given (up to a global shift) by
$$
\bet\equiv D_1\,R_1^{r_1-1}\,.
$$
The analysis of the Appendix shows that, for each $j\leq n-1$, the
level-$j$ strings $R_j$, $D_j$ are necessarily primitive.

\subsubsection{Two properties of admissible sequences}
To estimate the left hand side of \eqref{e:off-diag}, our first
objective is to count the number of admissible pairs
$j\stackrel{\vis,\bet}{\sim}j'$ (we recall that $j\sim j'$ implies
that $j\neq j'$). This counting is done in \S\ref{s:count} of the
appendix, and leads to the following
\begin{prop}\label{p:admissible}
There exists $C>0$ such that the following estimate holds. Fix
$\vis>0$. For any length $k>\vis$ and any primitive sequence $\bet$
of length $k$, the number of admissible pairs
$j\stackrel{\vis,\bet}{\sim} j'$ is bounded from above by $
C\,\vis^2 $.
\end{prop}
The number of terms in \eqref{e:off-diag} is thus uniformly bounded
when $k\to\infty$. It remains to control the variations of the
coefficients $c_{jm}(\bet)$ (defined in \eqref{e:normalization}),
and the size of $\cN(\bet)$. This is done in \S\ref{s:cjm} of the
appendix, and leads to
\begin{prop}\label{p:bounds}
Call $\Lambda=-\log|\lambda|=\frac12\log 3$.
For any $\vis$-admissible sequence $\bet$ of length $k>\vis$, any $m\in\{0,\ldots,k-1\}$
and any $\ell$, the coefficients $c_{jm}(\tau^\ell\bet)$ satisfy
\be\label{e:bound5}
-3\vis \Lambda\leq \log |c_{jm}(\tau^\ell\bet)| \leq 3\vis \Lambda\,,
\qquad j=1,\ldots,k\,.
\ee
\end{prop}
Notice that these bounds are not satisfied by all
$k$-sequences (see for instance the sequences used in
\S\ref{s:combination}). They are a consequence of the
almost-periodicity of $\bet$. These bounds straightforwardly imply
the following estimates for the normalization factor $\cN(\bet)$:
$$
|\lambda|^{6\vis}\,k\leq \cN(\bet)\leq |\lambda|^{-6\vis}\,k\,.
$$
Using Proposition~\ref{p:admissible} we get, for any
$\vis$-admissible $k$-sequence (and thus, trivially, for any primitive
$k$-sequence):
$$
\cN^{-1}\sum_{j\stackrel{\vis,\bet}{\sim}j'} |\overline{c_{jm}}\,c_{j' m}\,
\la \tau^{j}\bet|\Pi_{[\cdot\bep]_{\vis }}|\tau^{j'}\bet\ra|\leq
\cN^{-1}\sum_{j\stackrel{\vis,\bet}{\sim}j'} |c_{jm}\,c_{j' m}|\leq
C\,\vis^2\,|\lambda|^{-12\vis}\,k^{-1}\,.
$$
This ends the proof of Proposition~\ref{p:individual}, and thus of Theorem~\ref{th:3}.
$\hfill\square$

This Theorem strongly constrains the semiclassical measures one can
obtain from a family $(\Psi_{\bet}^m)_{k\to\infty}$, where
$\bet=\bet(k)$ and $m=m(k)$ are chosen arbitrarily. From
\cite{Rubin} we know that, if such a family converges to a
semiclassical measure $\rho$ with decay rate $e^{-\Gamma}$, then the
corresponding eigenvalues $|z_{\delta(k),m/k}|\to e^{-\Gamma/2}$,
which means that the relative degrees of the sequences $\bet(k)$
converge towards $\frac{\Gamma}{\log 3}$. The limit measure is then
of the form $\rho=\nu_{\Can}(dp)\times\nu(dq)$, with $\nu$ being the
limit of the measures $\nu_{[\bet(k)]}$. Although such limits $\nu$
can be quite diverse, they will necessarily satisfy the properties
of $\nu_{[\bet(k)]}$ described in Thm.~\ref{th:2}.

In the following section we exhibit semiclassical
measures which are {\it not} of the above type.

\section{Combination of two eigenstates $\Psi^m_{\bet}$}\label{s:combination}
In this section we prove Thm~\ref{th:4}, that is we provide examples
of semiclassical measures which are not of the form
$\nu_{\Can}(dp)\times\nu(dq)$. These measures will be associated
with  linear combinations of two particular degenerate eigenstates
$\Psi^m_{\bet}$, $\Psi^m_{\bet'}$.

Fix some complex number $z$ with $|\lambda|<|z|<1$. For any integer $k> 1$, we
can choose a degree $d=d(k)\in\{1,\ldots,k-1\}$ and $m=m(k)\in\{0,\ldots,
k-1\}$, such that the eigenvalues\be\label{e:limit}
z_{\delta(k),m/k}\stackrel{k\to\infty}{\longrightarrow} z\,,\quad\text{that
is,}\quad \frac{d(k)}{k}\to \delta(\infty)=\frac{\log |z|}{\log|\lambda|},\quad
\frac{m(k)}{k}\to \frac{\arg(z/\lambda^{\delta(\infty)})}{2\pi}\,. \ee For each
$k>4$, we then consider the two following $k$-sequences, which we choose to
label by indices $-k+d+1,\ldots,d$: \be\label{e:counterex}
\bet=\eta_{-k+d+1}\cdots\eta_d=(-)^{k-d}(+)^{d},\qquad
\bet'=\eta'_{-k+d+1}\cdots\eta'_d=(-)^{k-d -1}\,+-\,(+)^{d - 1}\,. \ee These
sequences have the same relative degree, and are primitive.
\begin{prop}
Consider the two eigenstates $\Psi^m_{\bet}$, $\Psi^m_{\bet'}$ constructed
from the sequences \eqref{e:counterex}, satisfying the condition \eqref{e:limit}.
Fix $\alpha,\,\alpha'\in \IC$ such that
$|\alpha|^2+|\alpha'|^2=1$.

Then, the sequence of eigenstates
$(\alpha\Psi^m_{\bet}+\alpha'\Psi^m_{\bet'})_{k\geq 1}$ converges to
a semiclassical measure $\mu_{\alpha,\alpha'}$. If
$\Im(\bar\alpha\alpha')\neq 0$, this measure is not of the type
$\nu_{\Can}(dp)\times\nu(dq)$.
\end{prop}
\begin{proof}
Let us first study the limit measure of the sequence $(\Psi^m_{\bet})$. One can
easily check that for $d\ge2\vis$ and $k-d\ge2v$ the sequence $\bet$ is not
$\vis$-admissible. Thus, from the results of the previous sections, the Husimi
weight of any $\vis$-square is given by \be\label{e:sum3}
H_{\Psi^{m}_{\bet}}([\bep'\cdot\bep]_{\vis})= \frac1{\cN}\sum_{j=-k+d + 1}^{d}
|c_{j0}|^2\,\la\tau^j\bet|\Pi_{[\bep'\cdot\bep]_{\vis}}|\tau^j\bet\ra =
\nu_{\Can}([\bep']_{\vis})\times\nu_{[\bet]}([\bep]_{\vis})\,. \ee
An explicit computation of the coefficients $c_{j0}(\bet)$ gives: 
\be\label{e:cj} \big(c_{j0}(\bet)\big)_{j=-k+d+1,\ldots,d} =\left(
\lambda^{(k-d-1)\delta},\ldots, \lambda^{2\delta},
\lambda^{\delta},1,\lambda^{(1-\delta)},\lambda^{2(1-\delta)},\ldots,
\lambda^{d(1-\delta)}\right)\,. \ee If we extend the sequence $\bet$ in
\eqref{e:counterex} to a bi-infinite sequence in ``the obvious way'' (that is,
taking $\eta_{j}=+$ for $j>d$ and $\eta_{j}=-$ for $j<-k+d+1$), and similarly
extend the coefficients $|c_{j0}|^2$ using the two geometric progressions, then
the extension of the sum in \eqref{e:sum3} to $j\in\IZ$ yields the weight of a
certain measure $\rho_{\delta(k)}=\nu_{\Can}\times\nu_{\delta(k)}$, which is an
exact eigenmeasure of $\cU$ of eigenvalue $|\lambda|^{2\delta(k)}$. Due to the
geometric decrease, the difference between the two measures is
small:
$$
\nu_{\delta(k)}([\bep]_{\vis}) =\nu_{[\bet]}([\bep]_{\vis}) +
\cO_{\vis}(|\lambda|^{\delta(1-\delta)k})\,,\qquad k\to\infty\,.
$$
In the limit $\delta(k)\to \delta(\infty)$, the measure
$\nu_{\delta(k)}$ converges to the eigenmeasure $\nu_{\delta(\infty)}$.

A similar computation shows that
the Husimi measure $H_{\Psi^m_{\bet'}}$ is close to an eigenmeasure
$\rho'_{\delta(k)} = \nu_{\Can}(dp)\times \nu'_{\delta(k)}(dq)$, which converges to
$\rho'_{\delta(\infty)}$ when $\delta(k)\to\delta(\infty)$.

Now, let us consider $\Psi_k \defeq \alpha\Psi^m_{\bet} +
\alpha'\Psi^m_{\bet}$, with $|\alpha|^2+|\alpha'|^2=1$. In the
equation
$$
H_{\Psi_k}([\bep'\cdot\bep]_{\vis})=|\alpha|^2\, H_{\Psi^m_{\bet}}([\bep'\cdot\bep]_{\vis})
+|\alpha'|^2\,H_{\Psi^m_{\bet'}}([\bep'\cdot\bep]_{\vis})
+2\Re \Big(\bar\alpha \alpha' \la \Psi^m_{\bet}|\Pi_{[\bep'\cdot\bep]_{\vis}}| \Psi^m_{\bet'}\ra\Big)\,,
$$
we need to control the cross-term, which is a linear combination of overlaps
$\la\tau^{j}\bet|\Pi_{[\bep'\cdot\bep]_{\vis}}|\tau^{j'}\bet'\ra$. From the
structures of $\bet$ and $\bet'$ this overlap is nonvanishing only if
$j=j'\in\{-\vis+1,\ldots,\vis-1\}$. Thus, the cross-term amounts to the finite
sum
$$
\frac{2}{\sqrt{\cN(\bet)\,\cN(\bet')}}\ \Re\Big(\bar\alpha \alpha' \sum_{j=-\vis+1}^{\vis-1}
\overline{c_{j0}(\bet)}\,c_{j0}(\bet') \la\tau^{j}\bet|
\Pi_{[\bep'\cdot\bep]_{\vis}}|\tau^{j}\bet'\ra\Big)\,.
$$
From the geometric decay of the coefficients $c_{j0}$, this sum takes the form
$\mu_{off,\delta(k)}([\bep'\cdot\bep]_{\vis}) +
\cO(|\lambda|^{\delta(1-\delta)k})$, where $\mu_{off,\delta}$ is a signed
measure (that is, the difference between two positive measures) which is
conditionally invariant under $\tcU$.
In the case of a square $[\eps'_1\cdot\eps_1]_1$, the above sum
reduces to a single term $j=0$:
$$
\overline{c_{00}(\bet)}\,c_{00}(\bet')\la f_{+}|\eps_{1}\ra
\la\eps_{1}|f_{-}\ra \la g_{-}|\eps'_{1}\ra \la\eps'_{1}|g_{+}\ra
=\begin{cases} -i/4\sqrt{3},&\ \eps_1=\eps'_1\in\{0,2\},\\
i/4\sqrt{3},&\ \eps_1\neq \eps'_1\in\{0,2\},\\
0&\ \text{otherwise}.
\end{cases}
$$
Thus, if $\Im(\bar\alpha\alpha')\neq 0$, we see that the signed measure
$\mu_{off,\delta}$ cannot be factorized into the form
$\nu_{\Can}(dp)\times\nu_{off,\delta}(dq)$. Hence, the semiclassical
measure $\mu_{\alpha,\alpha'}=|\alpha|^2\rho_{\delta(\infty)} +
|\alpha'|^2\rho'_{\delta(\infty)} + \mu_{off,\delta(\infty)}$ is not
of that form either.
\end{proof}

\section{Quantum unique ergodicity at the edges of the spectrum}\label{s:edge}

In the preceding sections we considered semiclassical measures with
eigenvalues in the ``bulk'' of the nontrivial spectrum, $|z|^2\in
(1/3,1)$. In this section, we restrict ourselves to eigenstates of
eigenvalues $z_{\delta,m}$ situated close to the {\it edges} of the
nontrivial spectrum, that is the circles $\{|z|=1\}$ and
$\{|z|=1/\sqrt{3}\}$. Since the analysis of the two cases are very
similar, we will mostly focus on the outer edge, that is the
vicinity of the unit circle. The eigenvalues $z_{\delta,m/\ell}$
satisfy $|z_{\delta,m/\ell}|=|\lambda|^{\delta}= 1+\cO(\delta)$, so
they will approach the unit circle iff the relative degrees
\be\label{e:outer-edge} \delta(k)\to 0\qquad\text{as}\quad
k\to\infty. \ee The general eigenstate of $z_{\delta,m/\ell}$ is a
linear combination of eigenstates $\Psi^m_{\bet}$ constructed from
sequences $\bet=\bet(k)$ of the same relative degree $\delta(k)$.
Notice that the periods $\ell(k)$ of $\bet(k)$ satisfy $\ell\geq
\delta^{-1}$, so they necessarily diverge when $k\to\infty$.

\subsection{Individual states $\Psi^{m}_{\bet}$ at the outer edge of the spectrum}\label{s:indiv}
As a first step towards the proof of Theorem~\ref{th:1}, we consider
the semiclassical measures associated with a family
$(\Psi^{m}_{\bet})_{k\to\infty}$ satisfying \eqref{e:outer-edge}.
From Proposition~\ref{p:individual}, we are reduced to studying the
limits of the associated measures $\nu_{[\bet]}$.
\begin{prop}\label{p:QUE-1}
Consider sequences $(\bet=\bet(k))_{k\to\infty}$ such that the
relative degrees $\delta(k)\to 0$ and their associated measures
$\nu_{[\bet(k)]}$ (see \eqref{e:factor}). Then, for any fixed
subinterval $[\eps_1\cdots\eps_{\vis}]$, we have
$$
\nu_{[\bet(k)]}([\bep])=\nu_{\Can}([\bep])+\cO_{\vis}(\delta(k))\,,
$$
where $\nu_{\Can}$ is the uniform measure on the Cantor set (see \eqref{e:nu_Can}).

As a consequence, the semiclassical measure associated with a family
$(\Psi^{m(k)}_{\bet(k)})_{k\to\infty}$
is $\rho_{max}=\nu_{\Can}(dp)\times\nu_{\Can}(dq)$.
\end{prop}
\begin{proof}
From the discussion in section \ref{s:husimi}, it is sufficient to prove the
proposition for primitive $k$-sequences $\bet$. A
sequence $\bet$ of relative degree $\delta(k)\ll 1$ will contain many more
minuses than pluses. It thus makes sense to split the sum in \eqref{e:factor}
between the indices $j$ such that the $\vis$-box of $\tau^j\bet$ contains {\em
only minuses}, and the indices $j$ for which the $\vis$-box contains at least
one plus. We write this decomposition as \be\label{e:decompo}
\nu_{[\bet]}([\bep])=\cN^{-1}\sum_{j}^{(-)}|c_{j0}|^{2}
\prod_{i=1}^{\vis}|\la\eps_i|f_{-}\ra|^2 +\cN^{-1}\sum_{j}^{(+)}|c_{j0}|^{2}\,
\prod_{i=1}^{\vis}|\la\eps_i|f_{\eta_{i+j}}\ra|^2\,. \ee Our aim is to show
that the second term on the right hand side becomes small when $k\to\infty$
and $\delta(k)\to 0$.
This will result from two facts. Firstly, since there are $\delta(k)k$ pluses
in $\bet$, the number of terms in $\sum^{(+)}$ is bounded from above by
$\vis\,\delta(k)\,k$, which is much smaller than the number of terms in
$\sum^{(-)}$ (larger than $k(1-\vis\,\delta(k))$).

Then, we also need to control precisely the variations of the coefficients
$|c_{j0}(\bet)|$ (which we will denote by $|c_{j}|$ for short).
These variations can be more easily visualized by considering the
logarithms
\be\label{e:B_j}
B_j(\bet)
\defeq \log|c_{j}(\bet)|=
\sum_{s=1}^{j}\log
\Big|\frac{\lambda_{\eta_s}}{\lambda^{\delta}}\Big|\,. \ee The
sequence $(B_j)_{j=0,\ldots,k}$ accomplishes a discrete path with
endpoints at the origin and two kinds of steps: \be\label{e:path}
B_{j+1}(\bet) - B_j(\bet) =\begin{cases}
 \delta \Lambda>0 & \text{if}\quad\eta_{j+1}=(-)\\
 (\delta-1) \Lambda<0 & \text{if}\quad\eta_{j+1}=(+)\,,
\end{cases}
\qquad \Lambda= \log|1/\lambda|\,. \ee For $\delta\ll 1$, the path will be made
of many small ups and few steep downs. Let us call $\{j_1<j_2<\ldots <j_d\}$
the indices such that $\eta_{j_r+1}=(+)$, and take $\ell_r=j_r- j_{r-1}$, so
that $\eta_{j_{r}+1}$ is preceded by a substring $(-)^{\ell_r-1}$. Grouping
together $|c_{j_r}|^2$ with the coefficients along the preceding substring, the
normalization factor can be written as \be\label{e:Cr} \cN(\bet)=
\sum_{r=1}^d C_r,\qquad C_r\defeq
|c_{j_r}|^2\sum_{m=0}^{\ell_r-1}|\lambda|^{2m\delta}=|c_{j_r}|^2\;
\frac{1-|\lambda|^{2\delta \ell_r}}{1-|\lambda|^{2\delta}}\,. \ee We now split
the above sum between the ``long'' and ``short'' $\ell_r$. We fix some
$\vareps\in (0,1/4)$ (independent of $\delta$), and consider the subsets of
indices
$$
\cL\defeq \{r\in [1,d]\ :\ \ell_r > \vareps/\delta\},\qquad \cS\defeq [1,d]\setminus \cL\,.
$$
One sees from (\ref{e:path}) that any index $j_{\max}$ at which $B_j$ reaches
its maximum is necessarily of the form $j_{\max}=j_r$ for some $r\in\cL$.
Conversely, for any $r\in\cL$, the coefficient $B_{j_r}$ is a ``local
maximum'', in the sense that $B_{j_r}>B_{j_{r}-1}$ and $B_{j_r}>B_{j_{r}+1}$.
We will show that the sum \eqref{e:Cr} is controlled by the ``long''
coefficients $j_r$:
\begin{lem}\label{l:cN2}
Consider the same assumptions as in Proposition~\ref{p:QUE-1}, and fix some $\vareps>0$.
Then there is a constant $C_\vareps>0$ such that, for $\delta$ small enough,
$$
C_\vareps^{-1}\,\delta^{-1}\sum_{r\in\cL} |c_{j_r}|^2\leq \cN(\bet) \leq
C_\vareps\,\delta^{-1}\sum_{r\in\cL} |c_{j_r}|^2\,.
$$
\end{lem}
\begin{proof}
We first estimate the contribution of ``long'' substrings to the sum
\eqref{e:Cr}: \be\label{e:cL}\forall r\in\cL,\qquad c'_{\vareps}\,\delta^{-1}\,
|c_{j_r}|^2\,\geq |c_{j_r}|^2\;\frac{1}{1-|\lambda|^{2\delta}} \geq C_r\geq
|c_{j_r}|^2\;\frac{1-|\lambda|^{2\vareps}}{1-|\lambda|^{2\delta}} \geq
c_{\vareps}\,\delta^{-1}\, |c_{j_r}|^2\,. \ee From (\ref{coeffs}) we have that
$|c_{j_r}|=|c_{j_{r-1}}||\lambda|^{1-\delta\ell_r}$. We then check that
\be\label{e:cS} \forall r\in \cS,\qquad C_r\leq
|c_{j_r}|^2\,\frac{1-|\lambda|^{2\vareps}}{1-|\lambda|^{2\delta}}\quad
\text{and}\quad
|\lambda|^{1-\delta}\leq \frac{|c_{j_r}|}{|c_{j_{r-1}}|}
\leq |\lambda|^{1-\vareps}\,. \ee The set of indices $\cS$ can be represented
as a disjoint union of ``discrete intervals'':
$$
\cS = \bigsqcup_{s} I_s ,\qquad\text{where  } I_s =
\{j\,|\,r_s\le j\le r_s+l_s-1\},\quad \text{and}\quad r_s+l_s<r_{s+1}\,.
$$ We are denoting by $l_s$ the length of the discrete interval $I_s$ and by $r_s$ its starting point.
Using the inequalities \eqref{e:cS}, we see that the contribution of
each interval $I_s$ to $\sum_r C_r$ is controlled by $j_{r_s-1}$,
which is the first ``long'' index at the left of $I_s$:
\be\label{e:cS-cL} \sum_{r\in I_s}  |c_{j_r}|^2 \leq
|c_{j_{r_s-1}}|^2 \; |\lambda|^{2(1-\vareps)}\frac{1- |\lambda|^{2(1-\vareps)l_s}}
{1-|\lambda|^{2(1-\vareps)}} \leq C\;|c_{j_{r_s-1}}|^2\,. \ee Taking
\eqref{e:cL} into account, we see that the sum \eqref{e:Cr} is of
the order of $\delta^{-1}\sum_{r\in\cL} |c_{j_r}|^2$.
\end{proof}
Any index $j$ in the sum $\Sigma^{(+)}$ is necessarily at distance $\leq \vis$
from some index $j_r$ (because the interval $[j-v,j+\vis]$ necessarily contains
a $(+)$), which implies $|c_{j}|\leq |\lambda|^{-\vis}|c_{j_r}|$. We thus get
$$
\sum_{j}^{(+)}|c_{j}|^2\leq C \sum_{r=1}^d |c_{j_r}|^2 \leq C'\,\sum_{r\in\cL} |c_{j_r}|^2\,.
$$
We used Eq.~\eqref{e:cS-cL} in the last inequality. Applying
Lemma~\ref{l:cN2}, we obtain the following upper bound for the
second sum in  \eqref{e:decompo}: \be\label{e:sum+}
\cN^{-1}\sum_{j}^{(+)}|c_{j}|^2= \cO(\delta),\qquad \delta\to 0\,.
\ee This implies the following estimate for the complementary sum:
$$
\cN^{-1}\sum_{j}^{(-)}|c_{j}|^2 = 1 -
\cN^{-1}\sum_{j}^{(+)}|c_{j}|^2 = 1+\cO(\delta)\,.
$$
The proof of the proposition is achieved by noticing that
$\prod_{i=1}^{\vis}|\la\eps_i|f_{-}\ra|^2 = \nu_{\Can}([\bep])$.
\end{proof}

\subsection{General eigenstates at the outer edge of the spectrum}
To prove the first part of Theorem~\ref{th:1},
we need to consider arbitrary
eigenstates, which are linear combinations of the states $\Psi^{m}_{\bet}$.

For $z$ inside the unit disk, we call $\cC_k(z)$ the set of orbits
$[\bet]$, such that $\bet$ is a $k$-sequence of relative degree
$\delta$, and there exists $m\in \{0,\ldots,\ell(\bet)-1\}$ such
that $z_{\delta,m/\ell}=z$ (we will only consider the case where
$\cC_k(z)$ is nonempty). We notice that the periods of two orbits
$[\bet],\ [\bet']\in\cC_k(z)$ may differ. On the other hand, to a
given orbit $[\bet]\in\cC_k(z)$ is associated a single integer $m\in
\{0,1,\ldots,\ell-1\}$ such that $z_{\delta,m/\ell}=z$. The states
$\{\Psi^{m}_{\bet},\ [\bet]\in \cC_k(z)\}$ form an orthonormal basis
of the $z$-eigenspace, so a general $z$-eigenstate will be written
$$
|\Psi\ra=\sum_{[\bet]\in \cC_k(z)} d_{\bet} |\Psi^{m}_{\bet}\ra\,,\qquad d_{\bet}\in\IC,\qquad
\sum_{[\bet]\in \cC_k(z)} |d_{\bet}|^2=1\,.
$$
For $k\gg\vis$, the Husimi measure of a vertical rectangle $[\cdot\bep]_{\vis}$ reads:
\be\label{e:Hus-general}
H_{\Psi}([\cdot\bep]_{\vis })=\sum_{[\bet],[\bet']\in \cC_k(z)}
\overline{d_{\bet'}}\, d_{\bet}\,
\la\Psi^{m'}_{\bet'}|\Pi_{[\cdot\bep]_{\vis }}|\Psi^{m}_{\bet}\ra\,.
\ee
The diagonal matrix elements can be
estimated using Proposition~\ref{p:QUE-1}:
\be\label{e:diag1}
\sum_{[\bet]\in \cC_k(z)} |d_{\bet}|^2
H_{\Psi^{m}_{\bet}}([\cdot\bep]_{\vis }) = \rho_{\max}([\cdot\bep]_{\vis })
+\cO_{\vis}(\delta)\,,
\ee
uniformly with respect to the normalized vector $(d_{\bet})$.

We now want to estimate the off-diagonal terms in
\eqref{e:Hus-general}. For two orbits $[\bet]\neq [\bet']$ in
$\cC_k(z)$, we will write $[\bet]\stackrel{\vis}{\sim}[\bet']$ if
there exists $(j,j')\in \IZ/\ell\IZ \times \IZ/{\ell'}\IZ$ such that
$\tau^j\bet$ and $\tau^{j'}\bet'$ coincide in the $\vis$-bulk. This
is possible only if the $\vis$-box of $\tau^j\bet$ contains some
pluses, and $\tau^{j'}\bet'$ consists in a reshuffling of these
pluses inside the box. For any $k$-orbit $\bet$, we call
$\reshuff(\bet)$ the set of $k$-sequences which coincide with $\bet$
in the $\vis$-bulk and have the same degree as $\bet$. Obviously,
$\# \reshuff(\bet)\leq \vis!$.

We define the following Hermitian matrix, indexed by the orbits $[\bet]\in \cC_k(z)$:
\be\label{e:matrix}
M_{[\bet'],[\bet]}\defeq
\begin{cases}
\la\Psi^{m'}_{\bet'}|\Pi_{[\cdot\bep]_{\vis }}|\Psi^m_{\bet}\ra\,,&\quad [\bet']\neq [\bet]\\
0\,,&\quad [\bet']= [\bet]\,.
\end{cases}
\ee
Observe that off-diagonal elements vanish unless $[\bet]\stackrel{\vis}{\sim}[\bet']$.
Our aim is to estimate the {\em spectral radius} of this matrix, $r_{\rm sp}(M)$.
If $\|v\|_{\infty}=\max_{[\bet]}|v_{[\bet]}|$ is the sup-norm in
the vector space of dimension $\#\cC_k(z)$, then the corresponding norm of the matrix
$M$ is given by
$$
\|M\|_{\infty} = \max_{[\bet]} M_{[\bet]},\qquad\text{where}\qquad
M_{[\bet]}=\sum_{[\bet']\in \cC_k(z)} | M_{[\bet],[\bet']}|\,.
$$
This norm $\|M\|_{\infty}$ is necessarily greater or equal to
the spectral radius $r_{\rm sp}(M)$.
For each $[\bet]$, the sum $M_{[\bet]}$ takes the form
\be
M_{[\bet]}=\sum_{[\bet']\stackrel{\vis}{\sim}[\bet]}
|\la \Psi^{m'}_{\bet'}|\Pi_{[\cdot\bep]_{\vis }}|\Psi^m_{\bet}\ra|\,.
\ee
Using the above remarks on the sequences $[\bet']\stackrel{\vis}{\sim}[\bet]$,
and calling $\{j_r\}$ the indices such that $\eta_{j_r+1}=(+)$, we find
$$
M_{[\bet]}\leq \cN(\bet)^{-1/2}\,\sum_{r=1}^d\sum_{j:|j-j_r|\leq \vis}
|c_{j}(\bet)| \sum_{\tbet\in \reshuff(\tau^j\bet)}
\cN(\tbet)^{-1/2}\,|c_0(\tbet)|\,.
$$
Any $\tbet\in \reshuff(\tau^j\bet)$ on the right hand side will belong to
the orbit $[\bet]$ or to some $[\bet']\stackrel{\vis}{\sim} [\bet]$.
Since $\tau^j\bet$ and $\tbet$ are identical outside the box, it is easy
to see that
$$
\tC_{\vis}^{-1}\leq \frac{\cN(\bet)^{-1/2}\,|c_{j}(\bet)|}{\cN(\tbet)^{-1/2}\,|c_{0}(\tbet)|}\leq
\tC_{\vis}
$$
for some uniform constant $\tC_{\vis}>0$. Using this estimate and Lemma~\ref{l:cN2},
we obtain the following upper bound:
$$
M_{[\bet]}\leq \vis!\,\tC_{\vis}\,\cN(\bet)^{-1}\sum_{r=1}^d\sum_{|j-j_r|\leq \vis} |c_{j}(\bet)|^2
\leq C'_{\vis}\; \cN(\bet)^{-1}\sum_{r=1}^d |c_{j_r}(\bet)|^2 = \cO_{\vis}(\delta)\,.
$$
This upper bound holds uniformly for all $[\bet]\in \cC_k(z)$, so it also applies
to $\|M\|_{\infty}$ and thus to $r_{\rm sp}(M)$. Since $M$ is Hermitian,
the off-diagonal part in \eqref{e:Hus-general} satisfies
$$
|d^\dagger M d |
= \Big|\sum_{[\bet'], [\bet]\in \cC_k(z)} \overline{d_{\bet'}}\,
M_{[\bet'],[\bet]} \, d_{\bet}\Big|
\leq r_{\rm sp}(M)\,\|d\|^2 = \cO(\delta)\,.
$$
This bound and \eqref{e:diag1} complete the proof of the first part of Theorem~\ref{th:1}
dealing with the outer edge of the spectrum.
\subsection{Inner edge of the spectrum}

The second part of Theorem~\ref{th:1} is proved in exactly the same way as the
first part, except that the sequences $\bet$ now consist of many $(+)$ and few
$(-)$. The Husimi measures of the corresponding eigenstates all converge to a
certain measure $\rho_{min}=\nu_{min}(dq)\times\nu_{\Can}(dp)$, where
$\nu_{min}$ is the self-similar measure defined as follows: \be\label{e:numin}
\forall\, \vis\text{-sequence }\bep,\qquad \nu_{min}([\bep])=\prod_{i=1}^{\vis}
|\la \eps_i | f_+ \ra|^2\,. \ee This measure is supported on the full interval,
so that $\supp\rho_{min}=K_+$. One easily checks that $\rho_{min}$ is
conditionally invariant through $\tcU$ with eigenvalue $1/3$. It is a Bernoulli
measure of the type considered in \cite{Rubin}. $\hfill\square$

\appendix
\section*{Appendix: $v$-admissible sequences}

We fix $\vis \geq 1$ and consider a primitive sequence $\bet$ of
length $k\gg \vis$, which is $\vis$-admissible. Our aim is to
analyze the structure of this sequence. We will proceed iteratively, from the
``macroscopic scale'' ($\sim k$) to the ``microscopic scale'' ($\sim
\vis$). At each step, one needs to consider several cases, so that
the set of possible structures can be represented by a ``tree''
organized into ``levels''. The structure of each admissible $\bet$
will correspond to a ``leaf'' of the tree situated at a certain
level $n$ (the sequence $\bet$ is then said to have ``rank $n$'').
Each rank-$n$ leaf will be characterized by a sequence of signed
integers \eqref{e:skeleton}. To fully specify $\bet$ (or rather its
orbit $[\bet]$), one further needs to give two ``elementary
strings'' $D_n$ and $R_n$. The construction of $\bet$ from these
data is explained in \S\ref{s:hierarchy}.

We now start to analyze $\bet$. We will present in detail the
analysis of the first two levels of $\bet$, and sketch the inductive
argument needed to get down to the ``microscopic'' level $n$. Our only
assumption is the existence of an admissible pair
$j\stackrel{\vis,\bet}{\sim} j'$. Up to a global shift of $\bet$, we
may assume that $j=0$ and $0<j'\leq k/2$. We designate $k_1=j'$ and
consider two cases, $k_1\leq\vis$ and $k_1>\vis$.

We recall the notation $\bet\equiv\bet'$ when
both sequences belong to the same orbit $[\bet]$; by $|\bet|$ we
denote the length of $\bet$. In all decompositions, curly brackets
$\{\cdots\}$ will indicate the part of the sequence lying in the
$\vis$-box.

\subsection{Case $k_1\leq \vis $: sequences of ``rank $1$''}\label{s:rank1}

\subsubsection{Structure of the sequence}
The assumption $\bet\sim\tau^{-k_1}\bet$ (the fact that the two sequences coincide in the $v$-bulk),
with $0 < k_1\leq \vis $, is equivalent to the following identity:
\be\label{e:condition-rank1} \eta_{\vis +1}\ldots\eta_{k}=\eta_{\vis
-k_1+1}\ldots\eta_{k-k_1}\,. \ee $i)$ If $\underline{k_1\geq
k-\vis}$, which is possible only when $k\leq 2\vis$, the index sets
 $\{\vis +1,\ldots,k\}$
and $\{\vis -k_1+1,\ldots,k-k_1\}$ do not overlap. The sequence $\bet$ can be written in terms
of two substrings $\bet^1$, $\bet^f$:
\be\label{e:short1}
\bet =\{ \bet^{f}\,\bet^1\} \,\eta^1_{1}\ldots\eta^1_{k-\vis}\,,\quad\text{with}\ \
|\bet^1|= k_1,\ \ |\bet^f|=\vis -k_1\,.
\ee
The two substrings can be chosen independently (as long as they satisfy the condition
that $\bet$ is primitive).

$ii)$ If instead we assume that $\underline{k_1< k-\vis}$, which
will be the case in the semiclassical limit, then the two index sets
in \eqref{e:condition-rank1} do overlap. If we call $\bet^1$ the
$k_1$-sequence $\bet^1=\eta_{\vis -k_1+1}\ldots\eta_{\vis }$, then
$\bet$ is constructed from a ``free'' initial part $\bet^f$ of
length $\vis -k_1$ and the repetition of $\bet^1$:
\be
\bet=\big\{ \bet^f\,\bet^1\big\} \,(\bet^1)^{q_1-1}\,(\eta^1_1\ldots\eta^1_{l_1})\label{e:short0}
\quad\Longrightarrow\quad \tau^{-l_1}\bet=\tbet^f\,(\bet^1)^{q_1}\,.
\ee
Here we have applied the Euclidean division $k-\vis
=k_1(q_1-1)+l_1$, with $0\leq l_1<k_1$, and set
$\tbet^f\defi\eta^1_1\ldots\eta^1_{l_1}\bet^f$, which has length
$<\vis $. In the nomenclature of \S\ref{s:hierarchy}, the sequence
\eqref{e:short0} has rank $1$, with elementary blocks $D_1=\tbet^f$,
$R_1=\bet^1$, and its structure reads $(r_1=q_1+1)$.

\underline{Remark:}
The string $\bet^1$ may not be primitive. Assume $\bet^1=(\tbet^1)^m$ for some $m\geq 1$,
with $\tbet^1$ primitive of length $\tk_1=k_1/m$.
Take $p,\,p'$ maximal such that $\tbet^f=(\tbet^1)^p\,\bet'\,(\tbet^1)^{p'}$, so
that
\be\label{e:prim1}
\bet\equiv \bet'\,(\tbet^1)^{\tq_1},\qquad \text{where}\quad \tq_1=mq_1+p+p'\,.
\ee
The ``defect'' $\bet'$ cannot be empty, otherwise $\bet$ would be periodic.

\subsubsection{Counting the admissible pairs $j\stackrel{\vis,\bet}{\sim} j'$\label{s:k1-counting}}
We remind that $j\stackrel{\vis,\bet}{\sim} j'$ means that $\tau^j\bet$ and $\tau^{j'}\bet$
coincide in the $\vis$-bulk $\{\vis+1,\ldots,k\}$. To estimate the number
of such pairs, we address the following question: knowing the orbit $[\bet]$
and the $\vis$-bulk of $\tau^j\bet$, what do we learn about $j$?

We separately consider the two cases (\ref{e:short1},\ref{e:short0}).

$i)$ The sequence \eqref{e:short1} has length $<2\vis$, so the number of
pairs is $<4\vis^2$.

$ii)$ Let us consider the sequence \eqref{e:short0}, or its
``irreducible form'' \eqref{e:prim1}. To estimate the number of
admissible pairs, we identify a (short) substring of $\bet$ which
allows us to identify the position of the defect along $\bet$.
\begin{lem}\label{l:defect1}
The string $\tbet^1\,\bet'\,\tbet^1$ occurs only once along the sequence
$\bet\equiv \bet'\,(\tbet^1)^{\tq_1}$. As a consequence, the string
$\bet^1\,\tbet^f\,\bet^1$ occurs only once as well.
\end{lem}
\begin{proof}
If $\tq_1=2$, the statement is equivalent with
the fact that $\bet$ is primitive. When $\tq_1\geq 3$, a fit of
$\tbet^1\,\bet'\,\tbet^1$ with a different substring of $\bet$
automatically implies that $\tbet^1$ is not primitive, which contradicts
our assumption.
\end{proof}
As a consequence, if the string $\bet^1\,\tbet^{f}\,\bet^1$ lies in
the $\vis$-bulk of $\tau^j\bet$, the shift $j$ can be uniquely
identified. On the other hand,
if $j\stackrel{\vis,\bet}{\sim} j'$ the string
$\bet^1\,\tbet^{f}\,\bet^1$ cannot be fully included in the
$\vis$-bulk of both partners, but it must intersect the $\vis$-box.
This string has length $\leq 3\vis$, so both indices $j,\,j'$ must
belong to the same interval of length $4\vis$. Hence, the total
number of admissible pairs $j\stackrel{\vis,\bet}{\sim} j'$ is less
than $16\vis ^2$.

\subsection{Case $k_1> \vis $}\label{s:higher-rank}

In this subsection we assume that $\vis <k_1\leq k/2$, then
decompose $k-\vis =(q_1-1)k_1+l_1$, with $0\leq l_1<k_1$. The
assumption $\bet\sim \tau^{-k_1}\bet$ is equivalent to
\be\label{e:condition0} \eta_{\vis +1}\ldots\eta_{k}=\eta_{k+\vis
-k_1+1}\ldots \eta_{k}\eta_1\ldots\eta_{k-k_1}\,. \ee This identity
implies that $\bet$ is determined by the subsequence
$\eta_{\vis +1}\ldots\eta_{\vis +k_1}$, which we baptize
$\bet^1=\eta^1_1\ldots \eta^1_{k_1}$: \be\label{e:long}
\bet=\big\{ \eta^1_{k_1-\vis +1}\ldots\eta^1_{k_1}\big\} \,
(\bet^1)^{q_1-1}\,(\eta^1_{1}\ldots\eta^1_{l_1})\,. \ee If $k_1$ and
$l_1+\vis $ were equal, we would have $k=q_1 k_1$, and the sequence
$\bet$ would be $k_1$-periodic, which is excluded by assumption.
Notice that $k_1$ is {\em strictly} smaller than $k/2$.

By inserting the above expression for $\bet$ into
\eqref{e:condition0}, we obtain a constraint on $\bet^1$:
\be\label{e:2d_case} \eta^1_{1}\ldots\eta^1_{k_1-\vis
}=\eta^1_{l_1+\vis+1}\ldots\eta^1_{l_1}\,,\quad
\text{equivalently}\quad (\bet^1)_i = (\tau^{l_1+\vis}\bet^1)_i,\ \
i=1,\ldots,k_1-\vis\,. \ee This constraint is similar with
\eqref{e:condition-rank1}. To compare the two situations, we also
need to know whether $\bet^1$ is primitive.
\begin{lem}\label{l:prim}
If $k_1<2\vis$ and $\bet^1=(\tbet^1)^m$ with $m>1$, where $\tbet^1$
is primitive of length $\tk_1=k_1/m<\vis$, then we are back to the
situation of \S\ref{s:rank1}: $\bet$ is of rank $1$, and there
exists an admissible pair $j\sim j'$ with $|j-j'|=\tk_1$.

If $k_1\geq 2\vis$, the string $\bet^1$ is necessarily primitive.
\end{lem}
\begin{proof}
Because $\bet$ is assumed primitive, we do not want $l_1+\vis$ to be
a period of $\bet^1$. If $\tk_1 \leq k_1- \vis$, the constraint
\eqref{e:2d_case} and the periodicity of $\bet^1$ imply that this
would be the case. In the opposite case $\tk_1 > k_1- \vis$, which
can occur only if $k_1<2\vis$, it is possible to realize the
constraint \eqref{e:2d_case} for $\bet^1=(\tbet^1)^{m}$, with
$l_1+\vis=m'\tk_1 + k_2$, $0<k_2<\tk_1$: this requires the identity
$\tbet^1_{1}\ldots \tbet^1_{k_1-\vis}=\tbet^1_{1+k_2}\ldots
\tbet^1_{k_1+k_2-\vis}$. In that case, $\bet\equiv
\tbet=(\tbet^1)^{\tq_1}\,\teta^1_1\ldots\teta^1_{k_2}\,$, which is
of the same form as in \eqref{e:prim1}, and forms an admissible pair
with $\tau^{-\tk_1}\tbet$.
\end{proof}
In the remainder of this
section we will assume that $\bet^1$ is primitive, and
separately consider the cases $k_1\gtrless \vis +l_1$.

\subsubsection{Case $k_1>\vis$ with  $\vis +l_1>k_1$}
We may write $\vis +l_1=k_1+k_2$, with, necessarily, $k_2<\vis $.

$i)$ In the case $k_2\geq k_1-\vis\defeq l_2$ (which can occur only
when $k_1< 2\vis$), we are in a situation similar to that of
\S\ref{s:rank1}, $i$: the condition \eqref{e:2d_case} does not
constrain $\bet^1$ very much, since the index sets
$\{1,\ldots,l_2\}$ and $\{1+k_2,\ldots,l_2+k_2\}$ do not overlap. In
that case, \be\label{e:rank2-0}
\bet^1=\bet^2\,\eta^2_{1}\ldots\eta^2_{l_2}
\,\bet^{f}=\bet^2\,\tbet^f\,,\quad\text{with}\ \ |\bet^2|=k_2,\ \
|\bet^{f}|=\vis-k_2\,, \ee the strings $\bet^2$, $\bet^f$ being
independent of one another. We must have $\tbet^f\neq \bet^2$,
otherwise $\bet$ would be $k_2$- periodic.

$ii)$ In the opposite case $k_2<k_1-v$, the situation is similar to
that in \S\ref{s:rank1}, $ii$. We divide $k_1-\vis =
(q_2-1)k_2+l_2$, $0\leq l_2<k_2$, $q_2\geq 2$. The constraint
\eqref{e:2d_case} implies that the sequence $\bet^1$ can be written
as \be\label{e:bet1-rank2}
\bet^1=(\bet^2)^{q_2}\,\eta^2_1\ldots\eta^2_{l_2}\,\bet^f=
(\bet^2)^{q_2}\,\tbet^f\,,\qquad |\bet^2|=k_2,\ \
|\bet^f|=\vis-k_2\,, \ee where the sequences $\bet^2$ and $\bet^f$
are independent. Notice that the sequence \eqref{e:rank2-0} has the
same form, with $q_2=1$. Inserting this expression in
\eqref{e:long}, we find
\begin{align}
\bet&=\big\{ \eta^2_{l_2+1}\ldots\eta^2_{l_2}\,\bet^f\big\} \,\Big((\bet^2)^{q_2}\,\tbet^f\Big)^{q_1-1}\,
(\bet^2)^{q_2}\,\eta^2_1\ldots\eta^2_{l_2}\,,\label{e:decompo5-0}\\
&\equiv\tbet^f\,(\bet^2)^{q_2+1}\,\Big(\tbet^f\,(\bet^2)^{q_2}\Big)^{q_1-1}
\label{e:tbet5}\,.
\end{align}
In the terminology of \S\ref{s:hierarchy}, this sequence is of
``rank 2'', with the structure $(+q_1,q_2+1)$, and the elementary
blocks $D_2=\tbet^f$, $R_2=\bet^2$

\medskip

The sequence $\bet^2$ is not necessarily primitive: it could be of
the form $\bet^2=(\tbet^2)^n$ with $\tbet^2$ primitive of length
$\tilde k_2$, and $n>1$. If we take $p,p'\geq 0$ maximal such that
$\tbet^f = (\tbet^2)^p\,\bet'\,(\tbet^2)^{p'}$, calling $\tq_2=n q_2
+p+p'$, we have: \be\label{e:primit} \bet\equiv
\bet'\,(\tbet^2)^{\tq_2+n}\,\Big(\bet'\,(\tbet^2)^{\tq_2}\Big)^{q_1-1}\,.
\ee Notice that $\bet'$ cannot be empty: it is a ``true defect''.
The following lemma is proven in a similar way to
Lemma~\ref{l:defect1}:
\begin{lem}\label{l:defect}
Assume $\bet^1$ is primitive.
Then the string $\tbet^2\,\bet'\,\tbet^2$ appears exactly $q_1$ times along $\bet$
of \eqref{e:primit}.
As a consequence,
the string $\bet^2\,\tbet^f\,\bet^2$ also appears $q_1$ times along $\bet$.
\end{lem}

\subsubsection{Case $k_1>\vis$ with $l_1 + \vis<k_1$}
In this case, the right hand side in the first equation of \eqref{e:2d_case} reads
$\eta^1_{l_1+\vis+1}\ldots\eta^1_{k_1}\eta^1_1\ldots\eta^1_{l_1}$.
We define
$$
k_2\defeq\min(l_1+\vis, k_1-(l_1+\vis))\,.
$$
In the three subcases below we will use the decomposition
$k_1-\vis= (q_2-1)k_2+l_2$, $0\leq l_2<k_2$.

\medskip

\hspace{0.5cm}\underline{Subcase $\vis +l_1=k_1-k_2$ with $0<k_2\leq \vis $}\\
In this case we have necessarily $q_1-1\geq 2$.
The condition \eqref{e:2d_case} implies that
\be
\bet^1=(\bet^2)^{q_2-1}\,\eta^2_{1}\ldots\eta^2_{l_2}\,\bet^f\,\bet^2
= (\bet^2)^{q_2-1}\,\tbet^f\,\bet^2\,,\qquad |\bet^2|=k_2,\ \ |\bet^f|=\vis -k_2\,.
\ee
Notice the similarity with \eqref{e:bet1-rank2}. The full sequence reads
\begin{align}
\bet&=\big\{ \bet^f\,\bet^2\big\} \,\Big( (\bet^2)^{q_2-1}\,\tbet^f\,\bet^2 \Big)^{q_1-1}\,
(\bet^2)^{q_2-2}\,\eta^2_{1}\ldots\eta^2_{l_2}\label{e:decompo4-0}\\
&\equiv
\big(\tbet^f\,(\bet^2)^{q_2-1}\big)\,\Big(\tbet^f\,(\bet^2)^{q_2}\Big)^{q_1-1}\,.\label{e:tbet4}
\end{align}
This sequence is of rank $2$, with the structure $(-q_1,q_2)$ and
the elementary blocks $D_2=\tbet^f$, $R_2=\bet^2$.
Lemma~\ref{l:defect} also applies here:
the string $\bet^2\,\tbet^f\,\bet^2$ occurs exactly $q_1$ times inside $\bet$.

\medskip

\hspace{0.5cm}\underline{Subcase $\vis +l_1=k_2 > \vis$}\\
From the condition \eqref{e:2d_case}, we may
write
\be\label{e:>2}
\bet^1 = (\bet^2)^{q_2-1}\,(\eta^2_{1}\ldots \eta^2_{l_2})\,
(\eta^2_{l_2+1}\ldots \eta^2_{l_2+\vis})\,,\qquad |\bet^2|=k_2\,.
\ee
$\bet^2$ satisfies some constraint of the form \eqref{e:2d_case}, depending
on $\vis+l_2\gtrless k_2$.

\medskip

\hspace{0.5cm}\underline{Subcase $\vis +l_1=k_1-k_2$ with $k_1/2\geq k_2> \vis $}\\
The condition \eqref{e:2d_case} imposes that $\bet^1$ can be expressed as
\be\label{e:>2bis}
\bet^1 = (\bet^2)^{q_2-1}\,(\eta^2_{1}\ldots \eta^2_{l_2})\,
(\eta^2_{k_2-\vis+1}\ldots \eta^2_{k_2})\,,\qquad
|\bet^2|=k_2\,.
\ee
$\bet^2$ satisfies some constraint of the form \eqref{e:2d_case}, depending
on $\vis+l_2\gtrless k_2$.

\subsection{Iterating the analysis}

In the last two subcases of \S\ref{s:higher-rank} ($k_1>\vis$ and
$k_2 > \vis $), the level-$2$ strings $\bet^2$ in \eqref{e:>2} or
\eqref{e:>2bis} satisfy constraints similar to \eqref{e:condition0}
(for $\bet$) \ or \eqref{e:2d_case} (for $\bet^1$). The analysis we
have performed successively on $\bet$ and $\bet^1$ can be applied to
$\bet^2$ and further iterated if necessary. At each step, we find
that the sequence $\bet^{j-1}\defeq R_{j-1}$ (of length $k_{j-1}$) is
composed of a ``repeated string'' $\bet^j\defeq R_j$ of length $k_j$, and
a ``defect'' $D_j$, as indicated in
(\ref{e:iteration1},\ref{e:iteration2}). This step determines the
signed integer $\sigma_j r_j$.

Since $k_{j}\leq k_{j-1}/2$, the lengths $k_1,k_2,\ldots$ decay
geometrically with $j$: for some $n\lesssim \log_2 k$, we end up
with a string $\bet^n=R_n$ of length $k_n\leq \vis $, and possibly
some extra string $\bet^f$ of length $<\vis$, which ends the
iteration. In general, the level-$n$ defect $\tbet^f=D_n$ is
obtained by adjoining to $\bet^f$ a strict substring of $\bet^n$.
$D_n$ and $R_n$ are the ``elementary strings'' of $\bet$. The latter
has rank $n$, structure $(\sigma_1r_1,\sigma_2r_2,\ldots,r_n)$, and can be
reconstructed from $D_n$, $R_n$ as explained in \S\ref{s:hierarchy}.

By applying Lemma~\ref{l:prim} at each step, we find that the
intermediate sequences $\bet^1,\ldots,\bet^{n-1}$ are primitive.
(The blocks $D_n$ and $R_n$ can be nonprimitive, see the remark
around \eqref{e:prim1} and the discussion around \eqref{e:primit}).

\subsubsection{Counting admissible pairs $j\stackrel{\vis,\bet}{\sim} j'$
for admissible sequences of rank $n$}
\label{s:count}

In this section we prove Prop.~\ref{p:admissible}, which estimates
the number of admissible pairs $j\stackrel{\vis,\bet}{\sim} j'$ for an arbitrary
$\vis$-admissible sequence $\bet$. This counting has been done
already for the sequences of rank $1$ in \S\ref{s:k1-counting}.
Below, the notation $\cS_\ell$ will stand for any of the two
level-$\ell$ strings $R_\ell,\,D_\ell$.

We give ourselves a sequence $\bet$ described by its structure $(\sigma_j\,r_j)$ and elementary
strings $D_n,\,R_n$.
We want to characterize the  admissible pairs $j\stackrel{\vis,\bet}{\sim}j'$,
that is, such that $\tau^j\bet$ and $\tau^{j'}\bet$ coincide outside the $\vis$-box.
In order to constrain those pairs, we will exhibit proper substrings of $\bet$ which
are ``identifiable'', or ``recognizable'' if they are contained in the $\vis$-bulk.
For instance, extending
Lemmas~\ref{l:defect1} and \ref{l:defect} to sequences of rank $n$, we see that
the string $R_n\,D_n\,R_n$ is recognizable. As a result, a defect $D_n$ can be
recognized if its ``neighbourhood'' $R_n\,D_n\,R_n$ is contained in the bulk.

The lower level strings $\cS_\ell$ can also be recognized if a certain ``neighbourhood''
lies in the bulk.
\begin{lem}
For any $\ell\leq n-1$ and any level-$\ell$ string $\cS_\ell=R_\ell/D_\ell$ of $\bet$,
we consider the following ``neighbourhood'' $\hat\cS_\ell$:
from the left end of $\cS_\ell$, take
$|R_n|$  steps on the left, and
$|S_\ell|+|S_{\ell+1}|+\ldots+|S_{n-1}| + 2|R_n|+|D_n|$ steps on the right
(here $S_i$ is the short level-$i$ string).

$\hat\cS_\ell$ automatically contains $\cS_\ell$.
If $\hat\cS_\ell$ is contained in the bulk, then the string $\cS_\ell$ it contains
can be recognized.
\end{lem}
A string which cannot be recognized is said to be ``hidden'' by the $\vis$-box.
\begin{proof}
Consider the level $n$: to recognize a string $\cS_{n-1}=D_{n}R_{n}^{r_n(-1)}$,
we need to see the defects $R_n\,D_n\,R_n$ adjacent to it,
that is, the bulk should contain the string
$R_n\,\cS_{n-1}\,D_n\,R_n$: from the left end of $\cS_{n-1}$, there
are $|R_n|$ steps on the left, and $|\cS_{n-1}|+|D_n|+|R_n|\leq
|S_{n-1}|+2|R_n|+|D_n|$ steps on the right.

In order to recognize $S_{n-2}$ (respectively $L_{n-2}$) we need to
identify $D_{n-1}\,R_{n-1}^{r_{n-1}-1}\,D_{n-1}$ (respectively
$D_{n-1}\,R_{n-1}^{r_{n-1}}$), therefore
$R_n\,S_{n-2}\,D_{n-1}\,D_n\,R_n$ (resp.
$R_n\,S_{n-2}\,R_{n-1}\,D_n\,R_n$) must be in the bulk. Whatever the
value of $\sigma_{n-1}$, the necessary distance on the right is at
most $|S_{n-2}|+|S_{n-1}|+2|R_n|+|D_n|$, while the distance on the
left is always $|R_n|$.

The proof for the lower levels proceeds by iteration.
\end{proof}
The identification of a level-$\ell$ sequence $\cS_\ell$ in the bulk of $\tau^j\bet$
implies that the same sequence can be identified at the same site in the bulk
of $\tau^{j'}\bet$.

If the level-$1$ defect $D_1$ were identifiable, we would have
$j=j'$, which contradicts the assumption $j\sim j'$. Thus its
neighbourhood $\hat D_1$ must intersect the $\vis$-box. This
provides a first restriction on $j,j'$.

To identify $D_1$, it would actually be sufficient to identify the
two strings $D_2$ adjacent to it. To avoid this, the box must
intersect one of the two neighbourhoods $\hat D_2$ adjacent to
$D_1$.

The lengths $|S_\ell|$
decay geometrically, $|S_{\ell+1}|< |S_{\ell}|/2$, so that $|\hat D_2|$
is bounded from above by
$|S_2|+2|S_3|+4\vis$. On the other hand, $|D_2R_2|=2|S_2|+|R_3|\geq 2|S_2|+|S_3|$.
Let us assume that $|S_2|>20\vis$. We then draw
$$
|D_2R_2| - |\hat{D}_2| \geq |S_2|- |S_3|-4\vis > |S_2|/2 -4\vis > 6\vis.
$$
As a result, the box can intersect at most a single one of the $r_1$
neighbourhoods $\hat{D}_2$, the other $r_1-1$ strings $\hat{D}_2$
sitting in the bulks of $\tau^{j}\bet$ and $\tau^{j'}\bet$. This
implies that
\begin{align*}
j'=j+k_1&\quad \text{if the hidden $D_2$ is on the left of $D_1$},\\
\text{respectively}\ \ j'=j- k_1&\quad \text{ if the hidden $D_2$ is on the right of $D_1$}.
\end{align*}
In the two cases, the two partners correspond to an exchange (a
``flip'') of two level-$1$ strings:
$$
R_1 D_1\to D_1 R_1,\quad\text{resp.}\quad D_1 R_1\to R_1D_1\,.
$$
Let us consider the first alternative ($j'=j+k_1$), and zoom on the string
$\hat D_2$ which intersects the box. Actually, to identify the $D_2$ it
contains, it would be sufficient to identify both level-$3$ strings $D_3$
adjacent to it. The box must thus intersect at least one of the neighborhoods
$\hat D_3$. Once more, if $|S_3|>20\vis$, only one of these neighborhoods can
be hidden. The choice of the hidden $D_3$ depends on $\sigma_1$. Assume for
instance $\sigma_1=-$, so that the defect $D_1=L_1=D_2\,R_2^{r_2}$. The flip
$R_1D_1\to D_1R_1$ then reads
$$
D_2\,R_2^{r_2-1}\,D_2\,R_2\,\,R_2^{r_2-1}\to
D_2\,R_2^{r_2-1}\,R_2\,D_2\,R_2^{r_2-1}\,,
$$
which involves the level-$2$ flip $D_2\,R_2\to R_2\,D_2$.
This shows that it is the string $D_3$ situated at the {\em
right} of $D_2$, that is the one at the junction $D_2\,R_2$, which
should be hidden. Iterating to higher levels, we see that, as long
as $k_\ell\gg\vis$, the exchange $\tau^j\bet\to \tau^{j+k_1}\bet$
involves either the flip $D_\ell\,R_\ell\to R_\ell\,D_\ell$ or the
opposite one, and the string $D_{\ell+1}$ at the junction must be
hidden: the box must intersect the corresponding neighborhood $\hat
D_{\ell+1}$. The iteration stops when $|S_{\ell}|\leq 20\vis$. At
this stage, the intersection of the box with $D_\ell$ implies that
$j$ must be contained in some interval of length $\leq 42\vis$
around the corresponding $D_{\ell}$. Since its partner $j'$ is
uniquely fixed by $j$, this proves the estimate in
Proposition~\ref{p:admissible}.
$\hfill\square$

\subsection{Variations of the coefficients $c_{jm}(\bet)$ for admissible sequences}\label{s:cjm}

In this section we will prove Proposition~\ref{p:bounds}, that is we show that,
for a sequence $\bet$ admitting partners $j\stackrel{\vis,\bet}{\sim} j'$, all
coefficients $|c_{jm}|$ are approximately of the same size.

\subsubsection{An alternative description of level-$\ell$ strings}
We will represent rank-$n$ admissible sequences $\bet$ in a slightly
different manner than in \S\ref{s:hierarchy}. Instead of
characterizing, at each level $\ell$, the strings $D_\ell$ and
$R_\ell$ by their lengths (``long'' vs. ``short''), we will rather
distinguish them by the relative number of elementary strings $R_n$,
$D_n$ they are composed of. That is, we will label differently the
branches and leaves of the tree representing the possible
admissible structures.

By convention, let us call ``positive'' (respectively ``negative'')
the elementary strings:
$$
P_n\defeq D_n=\tbet^f,\qquad N_n \defeq R_n=\bet^n\,.
$$
The two level-$(n-1)$ strings are now called as follows:
\be\label{e:level-n}
N_{n-1}=P_n\, N_n^{r_n},\quad P_{n-1}=P_n\, N_n^{r_n-1}\,.
\ee
Obviously, $N_{n-1}$ is the string containing more repetitions of $N_n$.

The construction of the lower levels proceeds by an iteration which
is different but similar to the one in
(\ref{e:iteration1},\ref{e:iteration2}). Starting from strings
$N_\ell$, $P_\ell$ at level $\ell<n$, we define a ``positive'' and a
``negative'' string at level $\ell-1$ by the following rule:
$P_{\ell-1}$ is the string with the highest number of $P_\ell$ or
the lowest number of $N_\ell$. The explicit form of $N_{\ell-1}$ and
$P_{\ell-1}$ depends on a signed integer $\varsigma_\ell r_\ell$,
where $r_\ell$ is the same as in \eqref{e:skeleton}:
\be\label{e:iteration}
\binom{N_{\ell-1}}{P_{\ell-1}}=\binom{N_{\ell}^{r_{\ell}}\,P_{\ell}}
{N_{\ell}^{r_{\ell}-1}\,P_{\ell}}\ (\varsigma_\ell=+)
\quad\text{vs}\quad
\binom{N_{\ell-1}}{P_{\ell-1}}=\binom{N_{\ell}\,P_{\ell}^{r_{\ell}-1}}
{N_{\ell}\,P_{\ell}^{r_{\ell}}}\ (\varsigma_\ell=-). \ee Except at
level $n$, we always place the $N_\ell$ to the left of the $P_\ell$,
so the above sequences are generally equal to $D_\ell$ or $R_\ell$
only up to appropriate shifts. The sign $\varsigma_\ell\in\{\pm\}$
indicates whether the  defect $D_\ell$ is (up to a shift) equal to
$P_\ell$ or $N_\ell$.
The string $N_\ell$ is a shift of either
$L_\ell$ or $S_\ell$, the choice depending on the signs
$\{\sigma_{n-1},\ldots,\sigma_{\ell+1}\}$, or equivalently
 $\{\varsigma_{n-1},\ldots,\varsigma_{\ell+1}\}$.

To be more synthetic, we call $P_\ell=\cS_\ell^+$ and $N_\ell=\cS_\ell^-$.
The iteration \eqref{e:iteration} means that the sequence $\cS_j^{\varsigma_j}$ is the
level-$j$ defect, while $\cS_j^{-\varsigma_j}$
is repeated $r_j$ or $r_j-1$ times in $\cS^\pm_{j-1}$.
The first integer ($-r_1$ vs. $+r_1$) corresponds to the global (lowest-level) structure of
$\bet$: for a certain shift $\tbet\equiv\bet$ one has
\be\label{e:level-1}
\tbet= N_{1}^{r_{1}-1}\,P_{1}\ (\varsigma_1=+)\quad \text{vs} \quad
\tbet= N_{1}\,P_{1}^{r_{1}-1}\ (\varsigma_1=-),\quad
\text{in short}\quad\tbet\equiv (\cS^{-\varsigma_1}_1)^{r_1-1}\,(\cS^{\varsigma_1}_1)\,.
\ee
We notice that $\bet$ contains more sequences $N_n=R_n$ than $P_n=D_n$.

\subsubsection{Variations of the $|c_{jm}(\bet)|$}
Let $\bet$ be the sequence described above, with relative degree $\delta=d/k$.
We first consider coefficients $|c_{jm}(\tbet)|$ associated with
the particular shift $\tbet$ of $\bet$ described in \eqref{e:level-1}.
The logarithms of the coefficients $|c_{jm}(\tbet)|$ (as in \eqref{e:B_j})
can be expressed in terms of a single ($\delta$-dependent) function
$$
\begin{aligned}
B:\bigsqcup_{n\geq 0}\{+,-\}^{n} & \longrightarrow \IR\\
\balpha=\alpha_1\cdots\alpha_n&\longmapsto
B(\balpha)=\sum_{s=1}^{n}\log \Big|\frac{\lambda_{\alpha_s}}{\lambda^{\delta}}\Big|\,,
\end{aligned}
$$
where we recall that $\lambda_-=1$, $\lambda_+=\lambda=i/\sqrt{3}$.
We then have $\log |c_{jm}(\tbet)|=B(\teta_1\cdots\teta_j)$.
As noticed in \S\ref{s:indiv}, these
coefficients form a discrete path made of a succession of ``ups''
$\delta\Lambda$ and `` downs'' $(\delta-1)\Lambda$, with $\Lambda=-\log|\lambda|$.

For any $n$-string $\bal$ we have the obvious bound \be\label{e:obvious}
|B(\alpha_1\cdots\alpha_n)|\leq n\,\Lambda\,. \ee In the
previous paragraph we have decomposed $\tbet$ into substrings,
starting at the highest level with the string $P_n$ which
initiates $\tbet$, and $N_n$ which follows it. We
renormalize the function $B$ by defining
$$
b(\bullet)\defeq \frac{B(\bullet)}{B(P_n)}.
$$
Equivalently, this function is defined as the unique function on
$\bigsqcup_{n\geq 0}\{+,-\}^{n}$, such that
$$
b(P_n)=1,\quad b(\tbet)=0\quad \text{and}\quad b(\balpha\bbeta)=b(\balpha)+b(\bbeta)\,.
$$
Since $|P_n|\leq\vis$, the bound \eqref{e:obvious} shows that $|B(P_n)|\leq \Lambda\vis$.
To prove Proposition~\ref{p:bounds} we will control the variations of the sequence
\be\label{e:sequ}
\{b(\tbet_1\cdots\tbet_n),\ 0\leq n\leq k\}\,.
\ee
Since $\tbet$ contains more strings
$N_n$ than $P_n$ and $b(\tbet)=0$, we have
$$
-1< b(N_n)<0<b(P_n)=1\,.
$$
Inspecting the alternative \eqref{e:iteration}, we see that at each level $1\leq \ell <n$, we have again
\be\label{e:order}
-1< b(N_\ell) < 0 < b(P_\ell) < 1\,.
\ee
This property reflects the name ``positive'' vs. ``negative''.
We can further constrain the values $b(N_\ell)$, $b(P_\ell)$.

Let us call $\#_\ell^{\pm}$ the number of
level-$\ell$ strings $\cS_\ell^{\pm}$ contained in the rank-$n$ sequence $\tbet$. The following
lemma relates this cardinal with the values of $b(\cS_\ell^{\pm})$.
\begin{lem}
There exists a real number $c>0$ such that, at each level $1\leq
\ell \leq n$, one has \be\label{e:identity}
b(N_\ell)=-\,c\,\#_{\ell}^{+},\qquad b(P_\ell)=
c\,\#_\ell^{-},\quad\text{or concisely}\quad b(\cS^{\pm}_\ell)=\pm
\,c\,\#_{\ell}^{\mp}\,. \ee The normalization condition $b(P_n)=1$
implies that $c=(\#_n^{-})^{-1}$.
\end{lem}
\begin{proof}
We reason by recurrence on increasing $\ell$.
From \eqref{e:level-1} we have at level $\ell=1$:
$$
0=b(\tbet)= (r_1-1)\,b(\cS^{-\varsigma_1}_1) +b(\cS^{\varsigma_1}_1)\,,\quad\text{and}\quad
\#_1^{-\varsigma_1}=r_1-1,\ \ \#_1^{\varsigma_1}=1\,.
$$
This means that there exists a real number $c$ such that
$$
b(\cS^{\varsigma_1}_1)=\varsigma_1\,c\,(r_1-1)= \varsigma_1\,c\,\#_1^{-\varsigma_1},
\qquad b(\cS^{-\varsigma_1}_1)=-\varsigma_1\,c=-\varsigma_1\,c\,\#_1^{\varsigma_1}\,.
$$
From \eqref{e:order} we must have $c>0$.
Let us now assume the property \eqref{e:identity} for some $\ell-1\geq 1$,
and first treat the case $\varsigma_\ell=+$,
so that the numbers of sequences of level $\ell$ are
$$
\begin{cases}
\#^+_\ell&=\#_{\ell-1}^{+}+\#_{\ell-1}^{-}\\
\#^-_\ell&= (r_\ell-1)\#_{\ell-1}^{+} + r_\ell\,\#_{\ell-1}^{-}\,.
\end{cases}
$$
At the same time, we easily extract the coefficients $b(\cS_\ell^\pm)$:
$$
\begin{cases}
r_\ell\,b(N_\ell)+b(P_\ell)&=-c\,\#_{\ell-1}^{+}\\
(r_\ell-1)\,b(N_\ell)+b(P_\ell)&=c\,\#_{\ell-1}^{-}
\end{cases}
\Longleftrightarrow
\begin{cases}
b(N_\ell)&=-c\,(\#_{\ell-1}^{+}+\#_{\ell-1}^{-})=-c\,\#^+_\ell\\
b(P_\ell)&=c\,((r_\ell-1)\#_{\ell-1}^{+} + r_\ell\,\#_{\ell-1}^{-})=c\,\#^-_\ell\,.
\end{cases}
$$
This proves the property at level $\ell$. The case $\varsigma_\ell=-$ is similar.
\end{proof}
\begin{lem}\label{l:sum}
Take $\tbet$ admissible of rank $n$. Then,
the values of $b$ on the defects $\cS^{\varsigma_\ell}_\ell$ satisfy:
$$
Sum(\tbet)\defeq \sum_{\ell=1}^{n-1} |b(\cS_\ell^{\varsigma_\ell})| =
\sum_{\ell=1}^{n-1} \varsigma_\ell\,b(\cS_\ell^{\varsigma_\ell}) < 1\,.
$$
\end{lem}
\begin{proof}
From \eqref{e:identity}, the above sum reads
$Sum(\tbet)=\frac{1}{\#_{n}^{-}}\sum_{\ell=1}^{n-1} \#_{\ell}^{-\varsigma_\ell}$.
On the other hand, if we call $\#_\ell=\#_\ell^+ + \#_\ell^-$ the total number of level-$\ell$ strings,
we check by recurrence that
$$
\forall \ell\leq n-1,\qquad \#_\ell = 1+\sum_{l=1}^{\ell} \#_{l}^{-\varsigma_l}\,.
$$
Indeed, we already have $\#_1=\#_1^{\varsigma_1}+\#_1^{-\varsigma_1}=1+ (r_1-1)$.
Assuming the above equality at level $\ell-1$,
the number of $\ell$-defects $\#_\ell^{\varsigma_\ell}$ is equal to the number $\#_{\ell-1}$ of
level-$(\ell-1)$ strings (one defect for each string), so that
$$
\#_{\ell-1}+\#_{\ell}^{-\varsigma_\ell}=\#_\ell^{\varsigma_\ell}+\#_\ell^{-\varsigma_\ell}=\#_\ell\,.
$$
This proves the recurrence. Thus, taking $\ell=n-1$ we get
$$
Sum(\tbet)=\frac{\#_{n-1}-1}{\#_{n}^{-}}=\frac{\#_{n}^{+}-1}{\#_{n}^{-}}\,.
$$
Finally, $\#_{n}^{+}< \#_{n}^{-}$ (these are respectively the numbers of strings $P_n$ and
$N_n$).
\end{proof}
We can now finish the proof of Proposition~\ref{p:bounds}. For any
level $1\leq \ell\leq n$, we call $b^\ell$ the ``sampling'' of the
sequence \eqref{e:sequ} obtained by keeping only the successions of
blocks of level $\ell$, starting from $b(\emptyset)=0,\ b(N_\ell)$,
and finally reaching $b(\tbet)=0$. The sequence $b^{\ell+1}$ is thus
a ``refinement'' of $b^\ell$.

We first describe the level $\ell=1$. If $\varsigma_1=-$, we have
$b^1\defeq \big(0,b(N_1),b(N_1P_1),\ldots,
b(N_1P_1^{r_1-1})=0\big)$. Its smallest value $b(N_1)$ is reached
after a ``steep drop'', then the sequence increases at a slower rate
to finally reach $0$ again. In the opposite case $\varsigma_1=+$,
the sequence $b^1=\big(0,b(N_1),b(N_1^2),\ldots,0\big)$ first slowly
decays until it reaches $b(N_1^{r_1-1})$, then it makes its largest
(positive) variation $b(P_1)$ to jump back to $0$. Its smallest
value is $b(N_1^{r_1-1})=-b(P_1)$. In both cases, the minimal
value of $b^1$ is $-|b(\cS_1^{\varsigma_1})|$.

Let us now study the variations of $b$ at the level $2<n$. First
assume $\varsigma_1=\varsigma_2=-$, so the sequence
$b^2=\big(0,b(N_2),b(N_2 P_2),\ldots,b(N_1),\ldots,0\big)$. It
first has a big negative jump $b(N_2)$, followed by $r_{2}-1$ small
positive jumps to reach $b(N_1)<0$, the smallest value of $b^1$.
Then starts the level-$2$ string composing $P_1=N_2 P_2^{r_2}$. From
$b(N_1)$ we have a steep negative jump to $b(N_1 N_2)$, then $r_2$
smaller positive jumps to reach $b(N_1 P_1)>b(N_1)$. The following
negative jumps in $b^2$ will never bring it as low as the value
$b(N_1 N_2)=b(N_1)+b(N_2)$, which is hence its smallest value. On
the other hand, all elements of $b^2$ (but the first and last) are
negative.

If $\varsigma_1=-$, $\varsigma_2=+$, we first have $r_2$ small
negative jumps to reach $b(N_2^{r_2})$, followed by a larger
positive jump of $b(P_2)$ to reach $b(N_1)<0$. Then, we have again
$r_2-1$ negative jumps to $b(N_1 N_2^{r_2-1})$, and a following
positive jump of $b(P_2)$ to get $b(N_1 P_1)>b(N_1)$. The following
values will consist of adding $b(P_1)$ to already existing values,
so they cannot get smaller. The smallest value of $b^2$ for this
case is thus $b(N_2^{r_2})=b(N_1)-b(P_2)$.

The case $\varsigma_1=+$ is equivalent to the ``time reversal'' of
the sequences $b^2$ described above. In all cases, the minimum of
$b^2$ occurs one step after or before the minimum of $b^1$, and its
value is given by
$$
\min b^2=\min b^1 - |b(\cS_2^{\varsigma_2})|
= -|b(\cS_1^{\varsigma_1})| -|b(\cS_2^{\varsigma_2})|\,.
$$
The reasoning can be pursued to find that at any level $\ell\leq
n-1$, the minimum of the sequence $b^\ell$ is given by $\min b^\ell=
- \sum_{l=1}^\ell|b(\cS_{l}^{\varsigma_{l}})|$, and that $b^\ell$
takes negative values except at its start and end.
At level $\ell=n-1$, we thus get $\min b^{n-1}=-Sum(\tbet)$. Once we
know $b^{n-1}=(0,b(N_{n-1}),\ldots)$, the sequence $b^n$ starts with
$b(P_n)=1$, followed by $r_n-1$ decays until it reaches
$b(N_{n-1})<0$. Since all values of $b^{n-1}$ are negative, we have
$b^{n-1}_i + b(P_n)\leq 1$ for any index $i$. On the other hand, the
value of $b^n$ never becomes smaller than $\min b^{n-1}$. As a
result, using Lemma~\ref{l:sum} we find that all the elements of
$b^n$ are bounded by
$$
-1 < -Sum(\tbet)\leq b^n_i\leq 1\,,\qquad 0\leq i\leq \#_n\,.
$$
By multiplying these inequalities by $B(P_n)$, we find that the components of
the rescaled sequence $B^n$ satisfy $|B^n_i|\leq \Lambda\vis$.
Finally, each string $\teta_1\ldots\teta_j$ is at most at ``distance'' $[\vis /2]$
from some string at level $n$, so using \eqref{e:obvious} we get the bound
$|B(\tbet_1\cdots\tbet_i)|=\log|c_{im}(\tbet)| \leq  3\Lambda \vis /2$
for any $0\leq i\leq k$.

Finally, the cocycle property
$c_{jm}(\tau^\ell\tbet)=\frac{c_{(j+\ell)m}(\tbet)}{c_{\ell m}(\tbet)}$
proves Proposition~\ref{p:bounds} for an arbitrary shift $\bet$
of $\tbet$.

$\hfill\square$


\end{document}